\documentstyle[preprint,aps,epsfig]{revtex}

\begin{document}

\draft
\preprint{
\vbox{
\hbox{ADP-99-9/T354}
}}

\title{Parton Distributions for the Octet and Decuplet Baryons} 

\normalsize
\author{C. Boros and A.W. Thomas}
\address {Department of Physics and Mathematical Physics,
                and Special Research Center for the
                Subatomic Structure of Matter,
                University of Adelaide,
                Adelaide 5005, Australia}

\date{\today}
\maketitle

\begin{abstract} 
We calculate the parton distributions for both polarized and unpolarized 
octet and decuplet baryons, using the MIT bag, dressed by mesons. We show that 
the hyperfine interaction 
responsible for the $\Delta$-$N$ and 
$\Sigma^0$-$\Lambda$ splittings leads to large deviations 
from $SU(3)$ and $SU(6)$  predictions.  
For the $\Lambda$ we find significant polarized, non-strange 
parton distributions which lead to a sizable 
$\Lambda$ polarization in polarized, semi-inclusive 
$ep$ scattering. 
We also discuss the flavour symmetry violation 
arising from the meson-cloud associated with the 
chiral structure of baryons.

\end{abstract} 

\tighten

\newpage 

\section{Introduction}

Parton distributions contain valuable information on 
the non-perturbative structure of hadrons.  
An impressive amount of data for both polarized and unpolarized structure 
functions on nucleon targets has been collected over the past 
two decades. However, relatively less is known about 
the parton  distributions in other baryons.   
Measurements of parton distributions for members of the baryon 
octet would give us complementary information to that obtained from
the nucleon and  
could shed light on many phenomena involving non-perturbative 
QCD, such as 
$SU(3)$ symmetry breaking, the flavor asymmetry in the nucleon sea 
and so on.  
   
Experimentally it should be  possible to access 
the parton distributions of $\Sigma^+$ hyperons 
through the Drell-Yan process. Furthermore, 
since the $\Sigma$'s are in general polarized 
because of their production mechanism, it should also 
be possible, in principle, to measure the 
polarized quark distributions in sigma hyperons.

It was recently pointed out by  Alberg {\it et al.} 
\cite{Alberg} 
that  the mechanism responsible for the splitting of the $\Delta$-$N$ and
$\Sigma^0$-$\Lambda$ masses could lead to considerable $SU(3)$
symmetry breaking in the parton  distributions  among members of
the baryon octet. Here, we show explicitly that 
this is indeed the case by calculating the quark distribution of baryons 
in the MIT bag model, where we include the  hyperfine  interaction which 
leads  to the splitting of the baryon masses.  
The $SU(3)$ breaking which we find goes beyond the implicit   
breaking of $SU(3)$  
by the strange quark mass, since it leads to 
deviations from $SU(3)$ expectations even among baryons 
with the same number of strange (valence) quarks. 
We also investigate the influence of  the meson cloud on the shape of the   
``bare''  quark distributions and calculate the flavor asymmetries 
in the sea arising from the meson-baryon fluctuations.

\section{Bare quark distributions} 

\subsection{Baryon Octet}

The starting point of our calculation is the general expression
for the quark distribution in a baryon $B$ with mass $m$
\cite{Jaffe83,Signal}:
\begin{equation}
   q_f(x)=\frac{m}{(2\pi)^3} \sum_n
   \int d^3p_n
        |\langle n; {\bf p}_n|\psi_+(0)| B\rangle |^2
      \delta [(1-x)m-p^+_n]
\label{eq1}
\end{equation}
Here $\psi_+ =\frac{1}{2} \gamma_- \gamma_+ \psi$
is the plus projection of the quark field
operator, the states $|n;{\bf p}_n\rangle $
are intermediate states with mass $m_n$ and form a
complete set of states with $p_n^+ = \sqrt{m_n^2+{\bf p}_n^2}
+p_{nz}$.
We stress that Eq. (\ref{eq1}) assures the correct support
for $q_f(x)$, regardless  of the approximations made for
$|n; {\bf p_n}\rangle$ and $|B \rangle$. 
The operator $\psi$ either destroys a quark 
in the initial state leaving a two quark system in the intermediate 
state  or it  creates an antiquark. Concentrating on the 
two quark intermediate states, and  
using MIT bag wave functions and the Peierls-Yoccoz method 
for constructing approximate momentum eigenstates,   
the spin-dependent parton distributions take the form 
\cite{Signal,Schreiber}  
\begin{equation}
q_f^{\uparrow\downarrow}(x) =
  \frac{m}{(2\pi )^2} \sum_{m} \langle B|P_{f,m}|B  \rangle
\int^\infty_{\frac{m^2(1-x)^2-m_n^2}{2m(1-x)}}
p_ndp_n \frac{|\phi_2({\bf p_n})|^2}{|\phi_3({\bf 0})|^2}
 |\Psi^{\uparrow\downarrow}_m({\bf p_n})|^2.
\label{eq:dist}
\end{equation}
Here $|\phi_2 ({\bf p_n})|^2$ and $|\phi_3({\bf 0})|^2|$ originate from the
Peierls-Yoccoz projections of the two quark intermediate
states and the (three quark) baryon, respectively.  
$ \langle B|P_{f,m}|B\rangle$
projects out the appropriate quantum numbers from the
spin-flavor wave function of the initial state.
$\Psi^{\uparrow\downarrow}_m({\bf p_n})$
are the Fourier transforms of the helicity and plus component 
projections,
$\Psi^{\uparrow\downarrow}({\bf x})=
\frac{1}{2}\gamma_- \gamma_+ \frac{1}{2}(1\pm\gamma_5 )\Psi_m({\bf x})$,
of the  MIT Bag wave function
\begin{equation}
 \Psi_m({\bf x}) =
N(\Omega ) \left(\sqrt{\frac{\omega +m_q}{\omega}}
 j_0(\frac{\Omega |{\bf x}|}{R}) \chi_m \atop {i
 \sqrt{\frac{\omega - m_q}{\omega}} ({\bf \sigma} . {\bf \hat{x}} )
j_1(\frac{\Omega |{\bf x}|}{R}) \chi_m } \right) 
\Theta (R-|{\bf x}|)
\end{equation}
with frequency $\omega =\sqrt{\Omega^2+(m_qR)^2}/R$, bag radius $R$ and
normalization constant $N(\Omega )$. $\Omega$ is the solution of the
eigenvalue equation
$tan( \Omega) =\Omega/(1-mR-\sqrt{\Omega^2+(mR)^2})$.

As we have already noted, 
the  advantage of using Eq.(\ref{eq1}) is that energy-momentum conservation 
is ensured and thus the  quark distributions obtained from it have correct 
support. \footnote{ 
Note that this is guaranteed by Eq.(\ref{eq1}) regardless
of the approximation used for the states $|n; {\bf p_n}\rangle$ and 
$|B\rangle$
-- in this case a Peierls-Yoccoz projection.}
The delta function implies 
that the distribution peaks at $x\approx (1-m_n/m)$,   
introducing a dependence of the shape of the quark distributions 
on the mass of the intermediate systems, $m_n$. 
Although intermediate states with higher number of 
quarks are possible, these contributions peak at negative 
$x$ values ($m_n >m$)  giving only a small contribution 
in the physical $x$-region.    
Thus, the main contribution for larger $x$ values comes from 
spectator systems with two quarks.  
Since the hyperfine interaction  
responsible for the splitting of the $\Delta$-$N$ masses 
also splits the masses of scalar and vector diquarks and 
whether the struck quark is accompanied by a scalar 
or vector diquark  is flavor dependent,    
this splitting  
leads to flavor dependent distortions in the shape of the quark distributions 
compared to exact $SU(6)$ symmetry. In the case of the nucleons, 
the $u$-quark distribution peaks at larger $x$-values than the 
$d$-quark distribution. 
These arguments for the explanation  of the observed $SU(6)$ 
violation in the quark distributions of the proton were 
first discussed in Ref. \cite{Close} and later implemented 
in the calculation of quark distributions 
in the MIT bag model for the proton \cite{Signal,Schreiber}. 
The same arguments  can be applied to other baryons.

It is instructive  to review the mass-splitting of the 
baryons here. The exact mechanism for this splitting is not essential 
for the calculation of the quark distribution since only the masses 
of the scalar and vector diquarks enter the calculation and  
different mechanisms/explanations lead to similar results. 
However, in order  
to illustrate how these numbers are obtained we discuss 
the one-gluon exchange model.  

The color hyperfine interaction Hamiltonian is given by
\begin{equation}
   H_{hf}=-\frac{1}{4} \sum_{i<j} v(m_i,m_j)\,\,
( \vec{\sigma}_i .  \vec{\sigma}_j)\,\, \lambda_i^a\lambda^a_{j}
\label{eq2}
\end{equation}
with $\frac{1}{2}\vec{\sigma}_i$ the spin of quark $i$ and $\lambda_i^a$ the
corresponding color matrix. The strength of the interaction
depends, in general, on the mass of the quarks. This dependence
is taken care of by $v(m_i,m_j)$ in Eq. (\ref{eq2}).
The sum over the color matrices can be calculated. One obtains
$-\frac{16}{3}$
for quark-antiquark pairs and $-\frac{8}{3}$
for baryons.
 
Attributing the entire mass splitting between the nucleon and the $\Delta$
to the hyperfine interaction, the splitting is given by
$\langle H_{hf}\rangle$. 
For three quarks, the spin sum in Eq. (\ref{eq2}) is 
$\sum_{i<j} ( \vec{\sigma}_i .  \vec{\sigma}_j)
    = (\vec{\sigma}_1 +\vec{\sigma}_2) .\vec{\sigma}_3
       + \vec{\sigma}_1.\vec{\sigma}_2 $.    
For a spin-0 and spin-1 diquark state, we have
$\langle \vec{\sigma}_1 .\vec{\sigma}_2\rangle_{S=0} =-3$  and
$\langle\vec{\sigma}_1 .\vec{\sigma}_2\rangle_{S=1} =1$, respectively.
Thus, one gluon exchange is attractive for scalar diquarks and
repulsive for vector diquarks.
Coupling the remaining
quark to the spin-triplet diquark state one obtains
$\langle \vec{\sigma}_3 . (\vec{\sigma}_1+\vec{\sigma}_2) \rangle =-4$  
for the nucleon and 
$\langle \vec{\sigma}_3 . (\vec{\sigma}_1+\vec{\sigma}_2)\rangle =2$ for the 
$\Delta$.    
Thus, the shifts in the nucleon and $\Delta$ masses are given by
  $\Delta m_N = -2  v(m_u,m_u)$   and
  $\Delta m_\Delta = 2 v(m_u,m_u)$, respectively and
the total splitting between the $\Delta$ and the nucleon is
$\Delta M = 4 v(m_u,m_u)$.  
Since $\Delta M$ is $\approx 300$ MeV 
we have $v(m_u,m_u)\approx 75$ MeV.  
The nucleon and the $\Delta$ would be
degenerate at $m=M_\Delta -2 v(m_u,m_u) \approx 
(1230 - 150)$ MeV $= 1080$ MeV, without hyperfine splitting.  
Further, we see that   the triplet diquark is heavier by $50$ MeV and
the singlet diquark is lighter by $150$ MeV than the diquark state
without hyperfine interaction.

The same arguments applied to the $\Lambda$ and $\Sigma$ lead to 
the following equations 
\begin{eqnarray}
 \Delta_{\Lambda} & = &\frac{2}{3}\, v(m_u,m_u)\, (-3) = -150 \mbox{MeV}
\nonumber\\
 \Delta_{\Sigma} & = &\frac{2}{3}\,
     \{v(m_u,m_u) - 4 v(m_u,m_s)\}.
\end{eqnarray}
Thus, for $v(m_u,m_s)$ we obtain with $m_\Sigma- m_\Lambda =
\Delta m_\Sigma-\Delta m_\Lambda$
\begin{equation}
  v(m_u,m_s)= v(m_u,m_u) -\frac{3}{8} (m_\Sigma -m_\Lambda) \approx 46
\mbox{MeV}. 
\end{equation}
$\Lambda$ and $\Sigma$ would be degenerate
with a mass of $\approx 1260$ MeV without hyperfine interactions.
The $us$  vector diquark
is heavier by $\frac{2}{3} v(m_u,m_s) \approx 30$ MeV
and the corresponding scalar diquark is
lighter by $2 v(m_u,m_s) \approx 90$ MeV than the diquark without
hyperfine splitting.

The mass of a diquark containing only $u$ and $d$ quarks
is about  $\frac{3}{4}$ of the degenerate mass of the nucleon and the
$\Delta$, which is roughly $800$ MeV. This gives us the masses
$m_s=650$ MeV  and $m_v=850$ MeV for
triplet and singlet diquarks containing $u$ and $d$ quarks.
To estimate the masses of diquarks containing a strange quark and
an up or down quark we use the phenomenological  
fact that the strange quark 
adds about $180$ MeV. Thus, we have $m_s^\prime=800+180-90\approx 890$ MeV
and $m_v^\prime=800+180+30 =1010$ MeV for singlet and triplet diquarks.

Having obtained the masses of the various diquark states, we turn our 
attention to  the quark distributions 
in different baryons.   One of the consequences of the mass differences 
between scalar and vector diquarks is that 
the up quark distribution in the proton peaks  at larger
$x$-values than the down quark distribution. 
To see this we note that  
the $SU(6)$ wave function of the proton is 
\begin{equation}
    p^\uparrow  = \frac{1}{3\sqrt{2}} 
 \left[ 3 u^\uparrow (ud)_{0,0}
   +  u^\uparrow (ud)_{1,0}
 -\sqrt{2} u^\downarrow (ud)_{1,1}  
 -  \sqrt{2} d^\uparrow (uu)_{1,0} + 2   d^\downarrow (uu)_{1,1} 
 \right] 
\label{eq:pwave}
\end{equation}
Here, we use the notation $(qq)_{S,S_z}$ for the diquark spin states 
with $S$ and $S_z$ the total spin and spin projection of the diquarks.  
While only  vector diquarks enter the calculation of 
the  down  quark distribution, both vector and scalar diquarks 
are relevant for the up quark distribution in the proton,  
with the scalar diquark having a much larger probability. 
According to the $\delta$ function in Eq. (\ref{eq1})    
the distribution of quarks accompanied by scalar diquarks, 
here the up quark distribution, 
should peak around  $x=1-m_s/m=1-650/940 \approx 0.31$ 
and that associated with a vector 
diquark, here the down quark distribution,  
around $x=1-m_v/m=1-850/940 \approx 0.1$ --- at the scale relevant for 
the bag model.   

The implementation is of these ideas is discussed in Ref.\cite{Schreiber} 
in detail. Here, we only note that the Fourier transform of the 
wave function in Eq. (\ref{eq:dist}) can be split into a spin dependent 
and a spin independent part
\begin{equation} 
   |\Psi_m^{\uparrow\downarrow}({\bf p_n})|^2 = 
  \frac{1}{2} \left[f({\bf p_n})\pm(-1)^{m+3/2} 
  g({\bf p_n}) \right].  
\end{equation} 
Since one uses a fixed polarization axis in the bag model  
the helicity states have to be projected out from the bag wave function
and thus  both polarization states, $m=\pm\frac{1}{2}$,  
contribute to a given helicity projection, $\uparrow$ or $\downarrow$.     
The expressions for $f({\bf p_n})$, $g({\bf p_n})$ and also 
for the Peierls-Yoccoz projections $|\phi_2({\bf p_n})|^2$ and 
$|\phi_3({\bf 0})|^2$ can be found in Ref. \cite{Schreiber} for massless 
quarks. The generalization to massive quarks 
is straightforward (multiplication of those parts of the expressions 
which comes from the upper and lower components by 
$\sqrt{(\omega \pm m_q)/\omega}$, respectively  
and using the normalization constant  
of the wave function for massive quarks).  

\begin{figure}
\psfig{figure=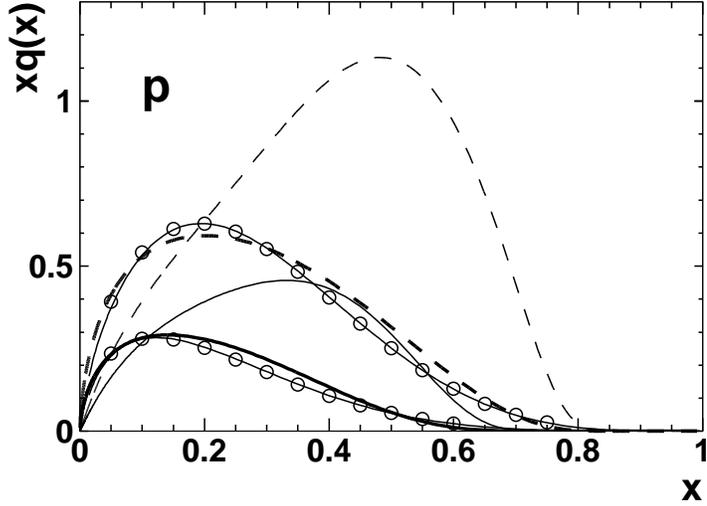,height=8.cm}
\caption{The up (dashed lines) and down 
(solid lines) valence quark distribution in the
proton 
at $Q^2=\mu^2=0.23$ GeV$^2$ (light lines)
and $Q^2=10$ GeV$^2$ (heavy lines). 
The quark distributions at $Q^2=10$ GeV$^2$ already include 
the meson-cloud corrections. 
The Cteq4M distributions representing the ``data''  
are shown as solid lines with open circles. } 
\label{fig1}
\end{figure}

Denoting by $F(x)$ and $G(x)$ those contributions to Eq. 
(\ref{eq:dist}) which come from the $f( {\bf p_n})$ and 
$g({\bf p_n})$ 
parts of the integral and using the 
the wave function of the proton (Eq. (\ref{eq:pwave}))  
to calculate the projections $\langle B | P_{f,m}|B \rangle$, 
we obtain 
\begin{eqnarray} 
  u^{\uparrow\downarrow}(x)& =& 
   \frac{1}{4}[F_v(x)+3F_s(x)]\mp\frac{1}{12}[G_v(x)-9G_s(x)],   
       \nonumber\\
   d^{\uparrow\downarrow}(x)& =& \frac{1}{2}F_v(x)\mp\frac{1}{6}G_v(x).
\label{eq:pdist}  
\end{eqnarray} 
Here, the subscripts, $s$ and $v$, on $F(x)$ and $G(x)$ indicate whether 
the intermediate states are scalar or vector diquarks. 
We calculated the quark distributions using $0.8$ fm for the 
bag radius; $m_v=850$ MeV and  $m_s=650$ MeV for the 
vector and scalar diquark masses. The result is shown 
in Fig. \ref{fig1} as light lines for a starting scale 
$\mu^2=0.23$ GeV$^2$.  
As discussed in Ref.\cite{Schreiber}, the two-quark intermediate 
states alone do not saturate the normalization of the quark distributions.   
There are also contributions from four quark intermediate states 
which have to be taken into account when normalizing the 
distributions.   
Since these peak at negative  
$x$-values due to the larger mass of the intermediate states,  
they give rise to distributions  which drop fast in the 
physical $x$-region.   
Here, we use the procedure adopted in the original paper 
\cite{Schreiber} and parametrize the four-particle contributions 
in the form $(1-x)^7$ (which gives an excellent approximation to the 
actual shape of the distributions)  such that 
the normalization is satisfied.   
After evolving the distributions to $Q^2=10$ GeV$^2$ (heavy lines) 
we find a good  agreement between the calculated  distributions and the 
experimental data which is represented by 
the CTEQ4M parametrization of the quark distributions \cite{Lai}. 
The results at $Q^2=10$ GeV$^2$ already contain corrections from 
the meson-cloud which will be discussed later. 
The quark distributions  
have been evolved in NLO with $\Lambda_{QCD} =0.225$ GeV and 
four active flavors, using the package of Ref. 
\cite{Kumano}.  

\begin{figure}
\psfig{figure=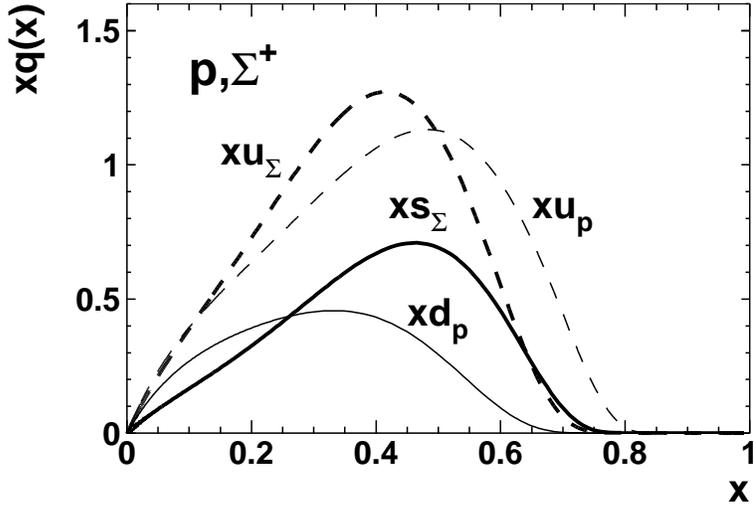,height=8.cm}
\caption{The strange (heavy solid line) and up 
(heavy dashed line) valence quark distributions in $\Sigma^+$ compared 
to the down (light solid line) and up (light dashed line) 
quark distributions in the proton 
 -- all evaluated at the bag scale, $\mu^2$. }
\label{fig2}
\end{figure}

\begin{figure}
\psfig{figure=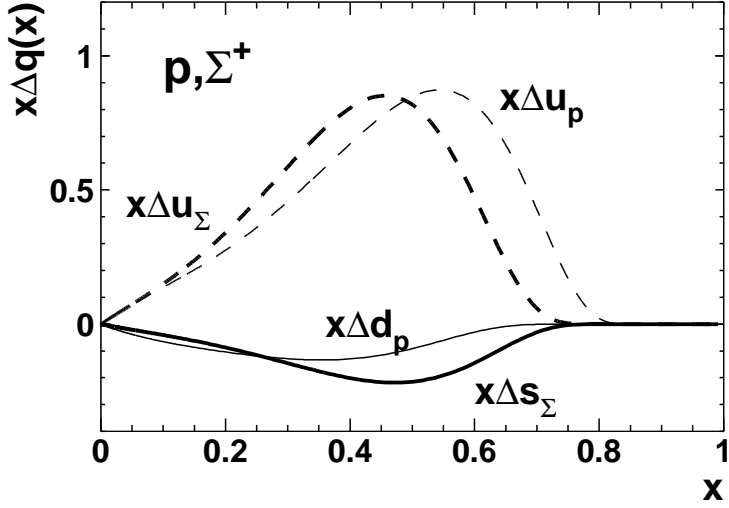,height=8.cm}
\caption{The 
polarized strange $x\Delta s(x)=xs^\uparrow (x) -xs^\downarrow (x)$  
(heavy solid line) and up $x\Delta u(x)$ 
(heavy dashed line) valence quark distributions in the $\Sigma^+$, compared 
to the polarized  down (light solid line) and up (light dashed line) 
quark distributions in the proton (at the bag scale, $\mu^2$). }
\label{fig3}
\end{figure}

Now, having fixed the parameters, let us generalize these 
arguments to the $\Sigma^+$. (This extension was first investigated 
semi-quantitatively in Ref.\cite{Alberg}.) The wave function of
the $\Sigma^+$ is given by Eq.(\ref{eq:pwave}) with the $d$-quark
replaced by an $s$-quark. (N.B. There must be a phase factor $-1$ relative 
to that of proton wave function, 
in order to match the phase convention 
of de Swart \cite{Swart} which we use.)     
The distribution of the strange quarks
is determined by the mass of the vector $uu$-diquark. It peaks at
$x=1-m_v/m_\Sigma=1-850/1190 \approx 0.29$, which is close to the value
found for the up quark distribution in the proton.
The maximum of the $u_\Sigma$ quark distribution
is determined  by the masses of both the $us$  scalar and vector  diquarks, 
which are  $\approx 890$ MeV and $1010$ MeV, respectively. 
Thus, $x=1-m^\prime_s/m_\Sigma=1-890/1190 \approx  0.25$ 
for the scalar diquark 
and  $x=1-m^\prime_v/m_\Sigma=1-1010/1190 \approx  0.15$ 
for the vector diquark,   
which are both  smaller than the corresponding values for  
$u_p$ and $s_\Sigma$. 
For the quark distributions we have, similar to Eq. (\ref{eq:pdist})
\begin{eqnarray} 
  u_{\Sigma^+}^{\uparrow\downarrow}(x)& =& 
   \frac{1}{4}[F^\prime_v(x)+3F^\prime_s(x)]\mp\frac{1}{12}[G^\prime_v(x)
    -9G^\prime_s(x)],   
       \nonumber\\
   s_{\Sigma^+}^{\uparrow\downarrow}(x)& =& 
  \frac{1}{2} F_v(x)\mp\frac{1}{6}G_v(x).
\label{eq:sigmadist}  
\end{eqnarray} 
Here, $F^\prime(x)$ and $G^\prime(x)$  differ from 
$F(x)$ and $G(x)$ because they 
are calculated by using 
the appropriate masses of the heavy diquarks and taking into account 
that one of the spectator  quarks is massive when making 
the Peierls-Yoccoz projections. In the calculation of $s(x)$ 
the struck quark is massive and the spectator quarks are 
massless. Thus, we calculate the Fourier transform 
$\Psi^{\uparrow\downarrow} ({\bf p_n})$ with the quark mass 
$m_q=180$ MeV and the Peierls-Yoccoz projections, 
$|\phi_2({\bf p_n})|^2$, with massless quarks. 
The results for the unpolarized and polarized distributions 
are shown in Fig. \ref{fig2} and Fig. \ref{fig3}    
(heavy lines)  together with 
the results for the proton (light lines).  
We see a considerable difference 
from the $SU(3)$ expectations, $s_\Sigma = d_p$ and 
$u_\Sigma = u_p$.   

\begin{figure}
\psfig{figure=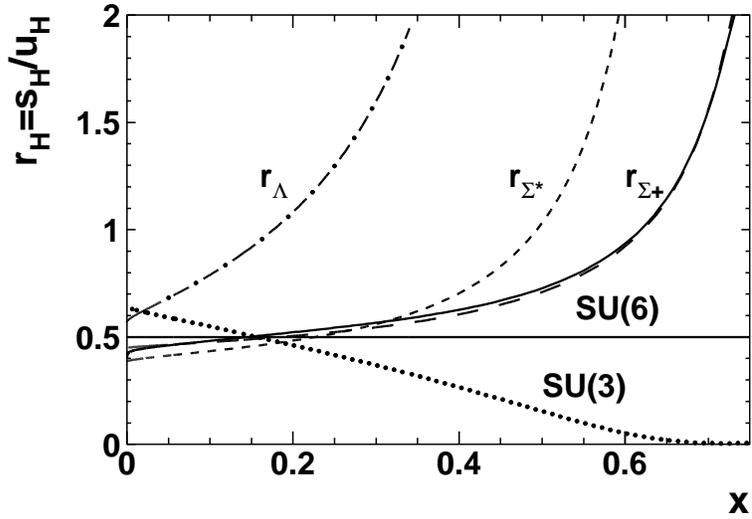,height=8.cm}
\caption{The ratios \protect $r_H\equiv s_{H}/u_{H}$  
for different baryons, after evolving the quark distributions 
to $Q^2=10$ GeV$^2$.  The ratio 
$r_{\Sigma^+}\equiv s_{\Sigma^+}/u_{\Sigma^+}$ is shown 
as the solid and dashed lines, with and without 
meson-cloud corrections, respectively. 
The $SU(3)$ expectation,   
which corresponds to $r_\Sigma =\frac{1}{2}r_\Lambda =r_p=d_p/u_p$,  
is shown as a dotted line.  
$SU(6)$ would give a constant ratio of $1/2$, independent of $x$ 
(solid line), and is realized for the decuplet baryons containing only 
massless  quarks  ($\Delta^+$). However, it is broken for 
the decuplet hyperons (short dashed line) -- see sec. IIB. }
\label{fig4}
\end{figure}

The  ratio $r_{\Sigma^=}\equiv s_{\Sigma^+}/u_{\Sigma^+}$ 
is shown in Fig. \ref{fig4} for both the bare quark distributions 
(solid line) and 
the distributions dressed by mesons (dashed line). 
We see that $r_\Sigma$ 
increases for $x\rightarrow 1$ in contrast to 
$SU(3)$ expectations which predict 
a behaviour  similar to that of $d/u$ in the proton.    
(The SU(3) expectation,  $r_{SU(3)}=r_p =d_p/u_p$, 
is  shown as the dotted line.) Exact $SU(6)$ symmetry would 
predict a constant ratio, independent of $x$, and this is shown as 
solid line in Fig. \ref{fig4}.    
We stress that these $SU(3)$ violations come partly through 
the explicit $SU(3)$ breaking by the strange quark mass and partly 
through the hyperfine interaction. $SU(3)$ breaking through the strange 
mass alone would not split the mass of the $\Lambda$ and $\Sigma$ 
hyperons and would lead to identical parton distributions 
in these hyperons. However, this is not the case and 
the hyperfine interaction plays a decisive role in the shape of the
parton distributions of the hyperons. 

The quark distributions of the  $\Lambda$ and $\Sigma^0$ hyperons
are interesting by themselves but we also need them  
to calculate the corrections arising from the  meson-cloud later. 
The $SU(6)$ wave function of the $\Sigma^0$
hyperon with given positive polarization is
\begin{eqnarray}    
    \Sigma^{0\uparrow } & = &
           \frac{1}{3\sqrt{2}} [  
    \sqrt{2} s^\uparrow (ud)_{1,0}   
    -2 s^\downarrow (ud)_{1,1}
  - \frac{1}{\sqrt{2}} d^\uparrow (us)_{1,0} + d^\downarrow (us)_{1,1}
   -\frac{3}{\sqrt{2}}d^\uparrow (us)_{0,0} + \nonumber \\
    & & -\frac{1}{\sqrt{2}}u^\uparrow (ds)_{1,0} + u^\downarrow (ds)_{1,1}
   - \frac{3}{\sqrt{2}}u^\uparrow (ds)_{0,0}].
\end{eqnarray}
The $ud$ diquark is always a vector diquark so that the 
maximum of the distribution 
of the strange quark is  determined only by the mass of the vector 
diquarks. Comparing with the wave function of the $\Sigma^+$ we see that  
$u_{\Sigma^0}=d_{\Sigma^0} = \frac{1}{2} u_{\Sigma^+}$ 
and $s_{\Sigma^0}=s_{\Sigma^+}$.  

\begin{figure}
\psfig{figure=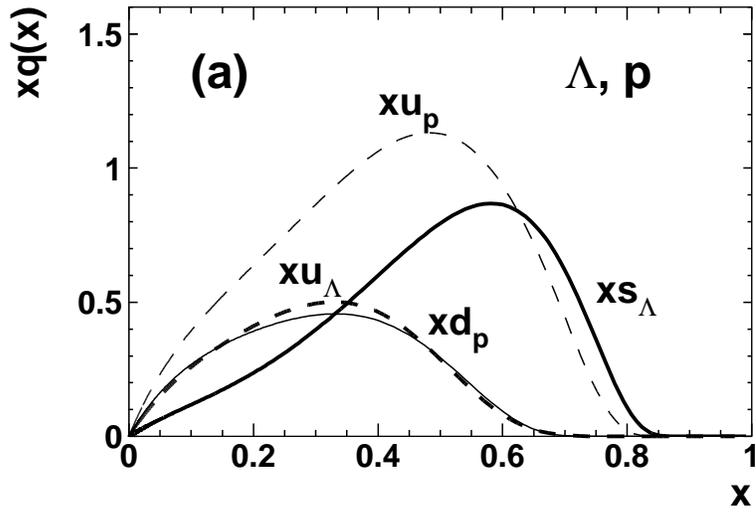,height=8.cm}

\psfig{figure=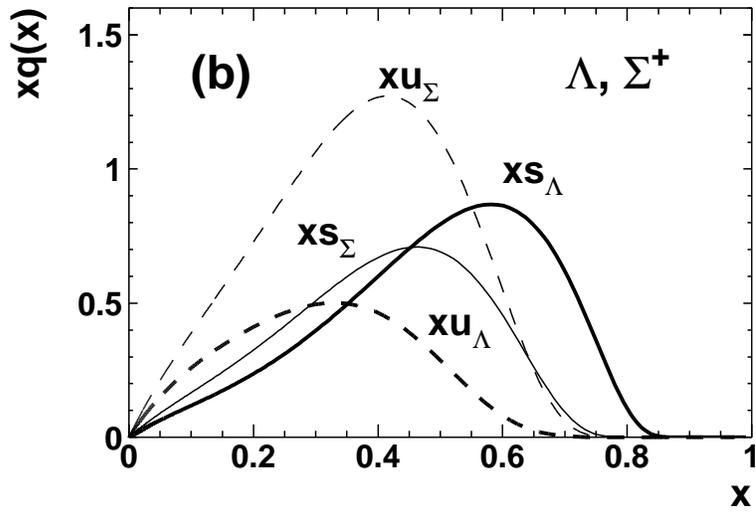,height=8.cm}
\caption{Quark distributions in the  $\Lambda$ 
compared to the quark distribution (a) in the proton  and (b) in  
the $\Sigma^+$ --  at the bag scale, $\mu^2$.}
\label{fig5}
\end{figure}

\begin{figure}
\psfig{figure=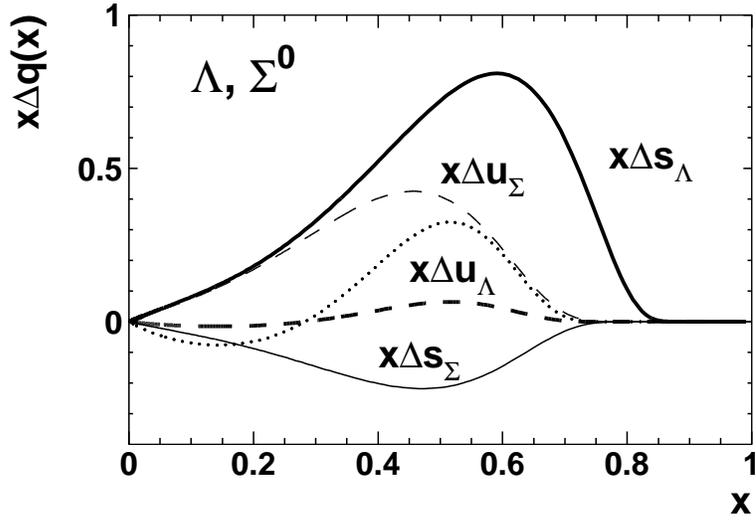,height=8.cm}
\caption{Polarized quark distributions in the  $\Sigma^0$ and the $\Lambda$ 
at the bag scale, $\mu^2$. 
The dotted line stands for five times $x\Delta u_\Lambda$ 
and indicates the relative importance of the $u$ and $d$ 
quarks in $g_1$.  }
\label{fig6}
\end{figure}

On the other hand,  
the $SU(6)$ wave function of the $\Lambda$ hyperon is  
\begin{eqnarray}
  \Lambda^\uparrow & = &
         \frac{1}{2\sqrt{3}}   [2 s^\uparrow  (ud)_{0,0}
+ \sqrt{2} d^\downarrow  (us)_{1,1}
 -d^\uparrow  (us)_{1,0} + d^\uparrow  (us)_{0,0} +\nonumber \\
  &&  -\sqrt{2} u^\downarrow   (ds)_{1,1}
 +u^\uparrow  (ds)_{1,0} - u^\uparrow    (ds)_{0,0} ]\,\,.
\end{eqnarray}
Whereas the maximum 
of the $u$ and $d$ distributions is determined by both
the vector and scalar diquark  masses,  only 
the mass of the scalar diquark is relevant for 
the maximum of the  distribution of the $s$-quark.   
$s_\Lambda$ peaks at $x=1-650/1115 \approx 0.42$. This yields 
a very hard distribution. For the $u$ and $d$ 
distributions we find that the peaks of the valence distributions
should occur around  
$x=1-890/1115 \approx 0.20$ and
$x=1-1010/1115 \approx 0.10$ for scalar and vector diquarks, respectively.  
The quark distributions of the $\Lambda^\uparrow$ are given by 
\begin{eqnarray} 
  u_\Lambda^{\uparrow\downarrow}(x)& =& 
   d_\Lambda^{\uparrow\downarrow}(x) =  
   \frac{1}{8}[3F^\prime_v(x)+F^\prime_s(x)]\mp\frac{1}{8}[G^\prime_v(x)
    -G^\prime_s(x)]  
       \nonumber\\
   s_\Lambda^{\uparrow\downarrow}(x)& =&\frac{1}{2}[ F_s(x)\pm G_s(x)].
\label{eq:lambdadist}  
\end{eqnarray} 
The results are shown in Fig. \ref{fig5}  and  Fig.  
\ref{fig6}  for the  polarized and unpolarized 
distributions  compared 
to the corresponding distribution in other baryons.  
The strange quark distribution in the $\Lambda$ is much harder 
than the corresponding strange quark distributions in the $\Sigma^+$ and 
$\Sigma^0$.  Large deviations from $SU(3)$  
expectations are most evident in Fig.\ref{fig4}, where the ratio 
$r_\Lambda\equiv s_\Lambda/u_\Lambda$,  shown 
as the dash-dotted line, is compared to the corresponding ratios in other 
hyperons. Exact $SU(6)$ would give $r_\Lambda =1$ and 
$SU(3)$ $r_\Lambda =2 r_p \equiv d_p/u_p$.

A naive approach to take into account the $SU(3)$ breaking would be 
to choose larger masses for strange quarks than for the up and down 
quarks and to argue that the strange quark distributions should 
peak at higher $x$-values than the light quark distributions 
due to its higher mass. Then, we would still  obtain 
$u_\Lambda =d_\Lambda = \frac{1}{2} u_\Sigma$ and $s_\Lambda =s_\Sigma$. 
However, this is not the case as can be seen in Figs.  
\ref{fig5} and  \ref{fig6}.   
We also see that,  
in contrast to the static quark model, the strange quark 
does not carry the total spin of the $\Lambda$ in the bag 
model, due to its transverse motion in the bag.  
Although the total contribution of the 
$u$ and $d$ quarks to the spin of the $\Lambda$  
(i.e. the integral over $\Delta u_\Lambda$ and $\Delta d_\Lambda$)   
is zero, the net polarization for given $x$  is non-vanishing. 
The splitting of the scalar and vector 
diquark masses shifts the light quark distributions 
with the same polarization as the $\Lambda$  
to higher $x$-values with respect to the 
corresponding distributions with opposite polarization.  
If  $G_v^\prime$ and $G_s^\prime$ 
had the same form,  
$\Delta u(x) =\frac{1}{2}(G_s^\prime (x) -G^\prime_v (x))$ 
would be zero.  
$\Delta u(x)$ and $\Delta d(x)$ are positive for large 
$x$ and negative for smaller $x$ values (see Fig. \ref{fig6}). 

It should be possible to test these results for the shapes of 
$\Delta u(x)$ and $\Delta d(x)$ in semi-inclusive deep inelastic scattering 
with longitudinally polarized electrons. 
Here, the smallness of the $u$ and $d$ polarizations relative 
to the strange quark polarization is compensated by the 
abundance of $u$-quarks in the valence region and by the fact 
that $s$ quarks are suppressed by a factor of $1/9$ compared to 
the corresponding factor of $4/9$ for the $u$-quark in electromagnetic 
interactions. (In Fig. \ref{fig6} we show five times $\Delta u(x)$ 
as a dotted line to indicate the relative magnitude of the 
contribution of $u$ and $d$ to $g_1^\Lambda$.)  
$\Lambda$'s produced in the current fragmentation region 
are mainly fragmentation products of $u$-quarks.  
Part of the polarization of the electron is 
transferred to the struck quark in the scattering process.  
This polarization will  be transferred to the final 
$\Lambda$ if the helicity dependent fragmentation 
functions, $\Delta D^\Lambda_u$  are non-zero \cite{Jaffe}. 
Since, according to the above discussion,  the 
$u$ and $d$ quarks in the $\Lambda$ hyperon may be  polarized 
at a fixed Bjorken $x$, we expect on general grounds 
that polarized 
$u$ and $d$-quarks may also fragment into a polarized $\Lambda$-hyperon. 
In fact, as pointed out by Gribov and Lipatov \cite{Gribov}, the 
fragmentation function $D^h_q(z)$, for a quark $q$ splitting 
into a hadron $h$ with longitudinal momentum 
fraction $z$, is related to the quark distribution 
$q_h(x)$, for finding the quark $q$ inside the hadron $h$ 
carrying a momentum fraction $x$, by the reciprocity relation 
\begin{equation} 
    D^h_q(z) \sim q_h(z) 
\label{eq:rec} 
\end{equation} 
for $z\sim 1$.   
Despite the limited range of validity of this relation, Eq.(\ref{eq:rec}) 
can serve as a first estimate of the fragmentation function 
\cite{Brodsky}. 
Since   $\Delta q^\Lambda_z$ is positive for large $x$ we expect 
to find positive polarization for $\Lambda$'s produced in the current 
fragmentation region. This is the opposite of the prediction of  
Jaffe \cite{Jaffe}, based on $SU(3)$ symmetry. 

\begin{figure}
\psfig{figure=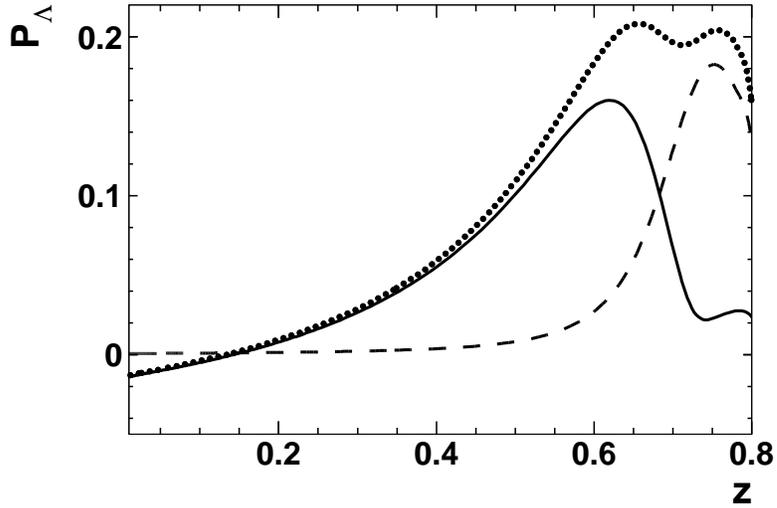,height=8.cm}
\caption{The polarization of the $\Lambda$ produced in semi-inclusive, 
polarized $e-p$ scattering, with the electron polarization arbitrarily 
set to 50\%. The contributions from 
the fragmentation of $u$ and $s$  quarks  are  shown  
as solid and dashed lines, respectively. The dotted line is the 
total polarization. } 
\label{fig7} 
\end{figure}

In order to estimate the expected $\Lambda$ polarization,  
we note that the polarization for the scattering of polarized electrons 
off an unpolarized target $N$ is given by \cite{Jaffe} 
\begin{equation} 
  \vec{P}_\Lambda = \hat{e}_3  P_e \frac{y(2-y)}{1+(1-y)^2}  
 \frac{\sum_q e_q^2 q_N(x,Q^2) \,\Delta D^\Lambda_q (z,Q^2) }
{\sum_q e^2_q q_N(x,Q^2)\, D^\Lambda_q (z,Q^2)},  
\end{equation} 
where $y\equiv (E-E^\prime)/E$ is the usual DIS variable; 
the electron beam defines the $\hat{e}_3$ axis   and 
$P_e$ is the degree of polarization of the incident electron.   
$P_\Lambda$ measures $\Delta D_u^\Lambda /D_u^\Lambda$ 
for not too small Bjorken-$x$ values, where the contributions from the 
strange quarks may be neglected. 
We calculated the $\Lambda$ polarization 
using $\Delta D^\Lambda_u = \Delta D^\Lambda_d$ and the 
reciprocity relation to replace the fragmentation functions by the 
quark distribution functions.   
The result calculated at   
$E_e\approx 30$ GeV, $x=0.3$ and $Q^2=10$ GeV$^2$, where 
$y=0.58$, is shown in Fig. \ref{fig7}. 
We assumed  a beam polarization of $50\%$.  
The solid and dashed lines are the contributions from the fragmentation of 
$u$-quarks and $s$-quark, respectively. The dotted line is the 
total polarization. The contribution of the $u$-quarks  dominates  
at $x\sim 0.5$. Since the $s$-quark distribution in $\Lambda$  
peaks at larger $x$-values  than the $u$-quark distribution, we also 
predict $D_u^\Lambda/D_s^\Lambda \rightarrow 0$ for $z\rightarrow 1$ for 
the fragmentation functions and, thus, 
the contribution of $s$-quarks to $P_\Lambda$ 
eventually dominates at very large $z$. However, since the 
cross section decreases rapidly with increasing $z$, the bulk of the 
produced $\Lambda$'s are fragmentation products of 
$u$-quarks. Thus, $P_\Lambda\ne 0$ at not too large $z$ 
will test our prediction.

\subsection{Baryon Decuplet}

Although the 
quark distributions of baryons from the baryon decuplet 
are unlikely to be measured in the near future they are 
of interest when we calculate the corrections associated with 
meson-baryon fluctuations. 

First of all let us check whether the values of $v(m_u,m_u)$ and 
$v(m_u,m_s)$ obtained from the $\Delta$-$N$ and  
$\Lambda^0$ and $\Sigma^0$ 
splittings are consistent with the values from the  
splitting of the $\Sigma$ and $\Sigma^*$ baryons.  
The masses of the $\Sigma^+$ and $\Sigma^{+*}$ are shifted by 
\begin{eqnarray} 
  \Delta m_{\Sigma^+} & = & \frac{2}{3}[ v(m_u,m_u) - 4 
 v(m_u,m_s)] \nonumber \\
  \Delta m_{\Sigma^{+*}} & = & \frac{2}{3}[ v(m_u,m_u) + 2 
 v(m_u,m_s)]   
\end{eqnarray}
with respect of the degenerate mass.    
Thus, the mass difference $m_{\Sigma^{+*}}-m_{\Sigma^{+}} 
= 4 v(m_u,m_s)$ gives $v(m_u,m_s) \approx 48$  MeV which is 
is very close to the value $v(m_u,m_s)\approx 46$ MeV obtained 
from the $\Lambda^0$-$\Sigma^0$ splitting.

Since the baryons in the decuplet are spin-3/2 particles, the 
spectator diquark system is always a vector diquark independent of 
the flavor of the struck quark and of the type of  baryon. 
This has the important consequence that the distributions 
of quarks of different flavor all have the same shape in the  
$\Delta$-baryons. Thus, SU(6) is a good symmetry for the 
$\Delta$ baryons. The distributions have a maximum 
at $x=1-m_v/m_\Delta\approx 0.31$ which is harder than the 
$d$-quark distribution in the proton, because of the larger 
mass of the $\Delta$, but somewhat softer than 
the distribution of the  $u$-quarks in the proton.  
Let us take the $\Delta^+$ as a representative for 
the $\Delta$ baryons and denote the spin projections 
$\pm \frac{1}{2}$ by $\uparrow\downarrow$ and 
$\pm \frac{3}{2}$ by $\Uparrow\Downarrow$. 
The $SU(6)$ wave function of $\Delta^{+\uparrow }$ may be written as 
\begin{equation} 
 \Delta^{+\uparrow } = \frac{1}{3} 
  \left[ d^\downarrow (uu)_{1,1}+\sqrt{2} d^\uparrow (uu)_{1,0} 
  + \sqrt{2} u^\downarrow (ud)_{1,1} +2 u^\uparrow (ud)_{1,0}\right] .  
\end{equation}   
The quark distributions 
of the $\Delta^{+\uparrow}$ are then given by 
\begin{eqnarray} 
   u_{\Delta^+}^{\uparrow\downarrow}(x) &=& 2 
      d_{\Delta^+}^{\uparrow\downarrow}(x)
       =  F_v(x)\pm \frac{1}{3} G_v(x) \nonumber\\
   u_{\Delta^+}^{\Uparrow\Downarrow}(x) &=& 2 
     d_{\Delta^+}^{\Uparrow\Downarrow}(x)
        = F_v(x)\pm  G_v(x) . 
\end{eqnarray}  

\begin{figure}
\psfig{figure=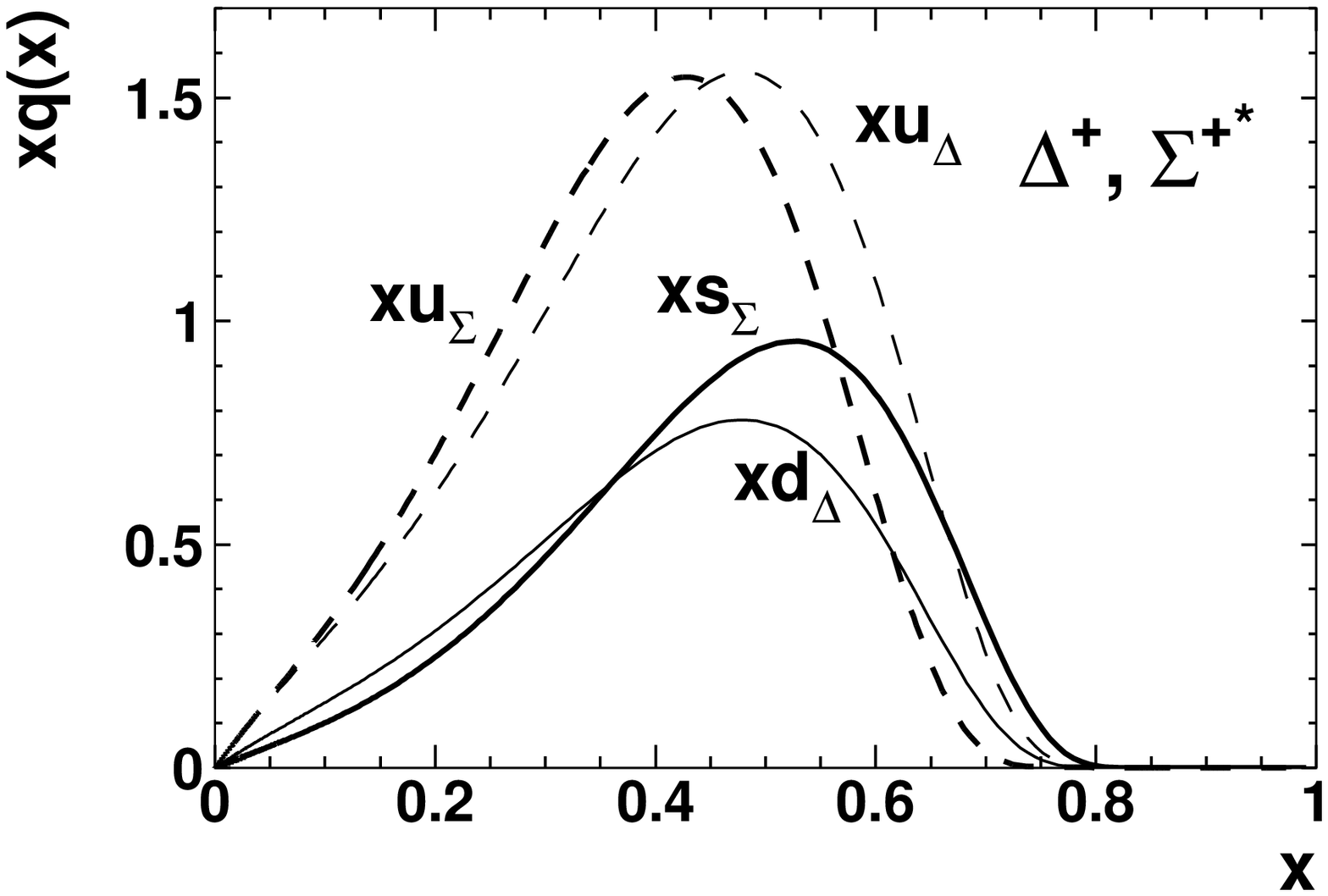,height=8.cm}
\caption{Quark distributions in the $\Delta^+$ and $\Sigma^{+*}$ 
at the bag scale, $\mu^2$.} 
\label{fig8}
\end{figure}

\begin{figure}
\psfig{figure=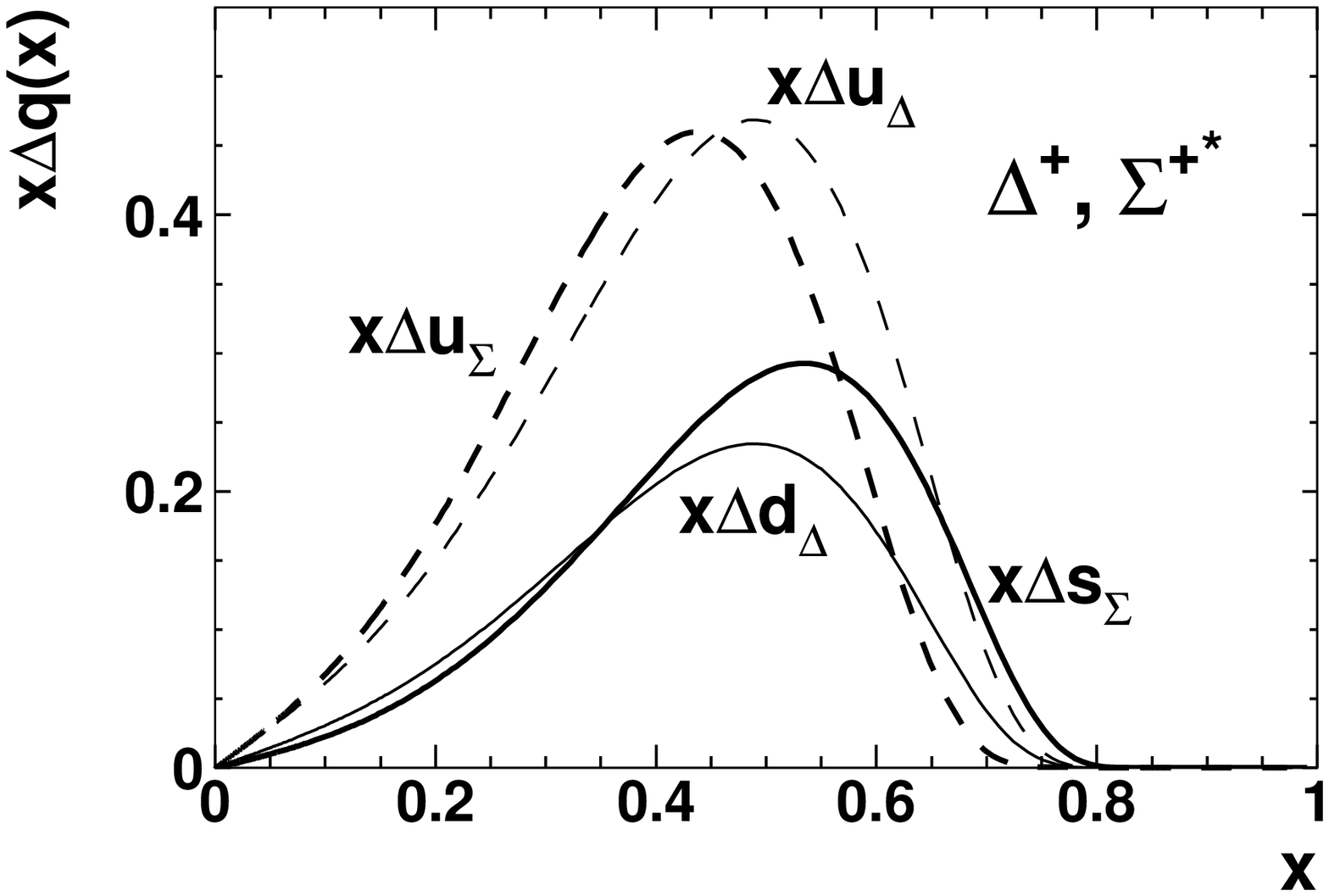,height=8.cm}
\caption{Polarized quark distributions in the $\Delta^+$ and $\Sigma^{+*}$ 
at the bag scale, $\mu^2$.  } 
\label{fig9}
\end{figure}

On the other hand, $SU(6)$ is broken for $\Sigma^*$. 
The up and/or down distributions in the   
$\Sigma^*$ baryons have a maximum at 
$x=1-m^\prime_v/m_{\Sigma^{*}}=1-1010/1385\approx 0.27$ and the 
strange quark distributions at 
$x=1-m_v/m_{\Sigma^*}=1-850/1385\approx 0.39$.  
The quark distributions, for example, for the  $\Sigma^{*+}$ 
are given by the same expressions as those for $\Delta^+$ 
replacing $d$ by $s$ and noting that the $u$ distribution is to be 
calculated with the heavy diquark masses and the $s$ distribution 
using the light diquark masses. 
Further, note that we have 
$\Delta q^{\frac{3}{2}}(x) \equiv q^\Uparrow (x) -q^\Downarrow (x) =  
3\Delta q^{\frac{1}{2}}(x) \equiv q^\uparrow (x) -q^\downarrow (x)$.   
In Figs. \ref{fig8} and  
\ref{fig9}  we show the unpolarized and polarized 
quark distributions in the $\Delta^+$ and 
$\Sigma^{+*}$. The $\Delta q$ are for the spin-$\frac{1}{2}$ 
projections. They have to be multiplied by $3$ to obtain 
the corresponding distributions for the spin-$\frac{3}{2}$ 
projections.  
In Fig. \ref{fig4}, we show the ratio 
$r_{\Sigma^*} \equiv s_{\Sigma^*}/u_{\Sigma^*}$ compared to 
the corresponding ratios in other hyperons.

\section{Meson cloud corrections} 

The importance of  the chiral structure of nucleons 
is well established both experimentally and theoretically.   
The pion-cloud associated with chiral symmetry breaking 
was first discussed in the context of deep-inelastic scattering 
by Feynman \cite{Feynman} and Sullivan \cite{Sullivan}. 
It  leads to  flavor symmetry violation (FSV) 
in the sea-quark distributions  of the nucleons,  
as realized by Thomas \cite{Tho83}.  
FSV in the proton was first observed experimentally by 
the NMC Collaboration through a violation of the 
Gottfried sum-rule \cite{NMC}. More recently it has been directly 
studied by the NA51 Collaboration at CERN  \cite{NA51},  
by the E866 Collaboration at Fermilab \cite{E866} and 
by the Hermes Collaboration at Desy \cite{HER}.   
The role of the meson-cloud in understanding these data  
have been discussed extensively in the literature  \cite{PiGott,MT98}  
--- for recent reviews see Ref \cite{SpeT,LonT,Kum}.  
On the other hand, relatively less  
attention has been  paid to  the effects of the meson-cloud 
in other baryons.  As pointed out by Alberg {\it et al.},  
and discussed in more detail in Ref. \cite{Alb98}, 
the meson-cloud predicts an excess of $\bar d$ over 
$\bar u$ in $\Sigma^+$ hyperons similar to that in observed in protons,  
while SU(3) suggests  $\bar d < \bar u$, since under 
$p\leftrightarrow \Sigma^+$ we have $s(\bar s) \leftrightarrow d(\bar d)$.  
The meson-cloud also modifies 
the bare quark distributions of the hyperons.   
In the following we discuss both  FSV in hyperons 
and the modification of the bare quark distributions due to 
the meson-cloud. 

In order to take account of the chiral structure of a baryon, its 
wave function is written 
as the sum of meson baryon Fock states 
\begin{equation} 
   |H \rangle = \sqrt{Z}|H\rangle_{bare} 
+\sum_{BM}\int dy d^2\vec{k}_\perp \phi_{BM}(y,k_\perp^2) 
 |B(y,\vec{k}_\perp );M(1-y,-\vec{k}_\perp ) \rangle . 
\end{equation} 
Here $\phi_{BM}(y,k_\perp^2)$ is the probability amplitude for the 
hyperon to fluctuate into a virtual baryon-meson $BM$ system with 
the baryon and meson having 
longitudinal momentum fractions $y$ and $1-y$ and transverse momenta 
$\vec{k}_\perp$ and $-\vec{k}_\perp$, respectively. 
$Z$ is the wave function renormalization constant and is equal to 
the probability to find the bare hyperon in the physical hyperon.   

In the following we discuss the chiral structure 
of the $\Sigma^+$ as an example and compare it to that of the 
nucleons. The nucleon case has already been discussed in   
\cite{Steff}. The extension to other baryons 
is straightforward. 
The lowest lying fluctuations for $\Sigma^+$ which we include 
in our calculation are 
\begin{eqnarray}
    \Sigma^+(uus)& \rightarrow  &\Lambda^0(uds)\,\pi^+(u\bar d) \nonumber \\
    \Sigma^+(uus)& \rightarrow  &\Sigma^0(uds)\,\pi^+(u\bar d) \nonumber \\
    \Sigma^+(uus)& \rightarrow  &\Sigma^+(uus)\pi^0(
  \frac{1}{\sqrt{2}} [d\bar d - u\bar u]) \nonumber \\
    \Sigma^+(uus)& \rightarrow  &\Sigma^{0*}(uds)\,\pi^+(u\bar d) \nonumber \\
    \Sigma^+(uus)& \rightarrow  &\Sigma^{+*}(uus)\,\pi^0(
  \frac{1}{\sqrt{2}} [d\bar d - u\bar u]) \nonumber \\
     \Sigma^+(uus)& \rightarrow  &p(uud)\, \bar K^0(\bar d s). 
\label{sfluct}
\end{eqnarray}
The corresponding lowest fluctuations 
for the proton are 
\begin{eqnarray} 
     p (uud) &\rightarrow  & n(udd) \,\pi^+(u\bar d)\nonumber\\ 
     p (uud) &\rightarrow  & p(uud)\,  
 \pi^0 ( \frac{1}{\sqrt{2}} [d\bar d - u\bar u])\nonumber\\ 
     p (uud) &\rightarrow  & \Delta^+(uud) 
 \,\pi^0 (\frac{1}{\sqrt{2}} [d\bar d - u\bar u])\nonumber\\ 
  p (uud) &\rightarrow  & \Delta^{0} (udd) \,\pi^+(u \bar d)\nonumber\\ 
  p (uud) &\rightarrow  & \Delta^{++} (uuu) \,\pi^-(\bar u d).   
\end{eqnarray}      
Since the $\Delta$ plays an important 
role in the nucleon, we also include the $\Sigma^*\pi$ components of the 
wave function in the $\Sigma^+$ case.

In deep inelastic scattering, the virtual photon can hit either  
the bare hadron, $H$, or one of the constituents of the higher 
Fock states.  
In the infinite momentum frame, where the constituents of the target  
can be regarded as free during the interaction time, the contribution 
of the higher Fock states to the quark distribution of the 
physical hadron, $H$,  can be written as the convolution 
\begin{equation}
 \delta q_{H}(x) = \sum_{MB} \left [
  \int_x^1 f_{MB/H}(y) q_M (\frac{x}{y})
\frac{dy}{y} +
  \int_x^1 f_{BM/H}(y) q_B (\frac{x}{y})
\frac{dy}{y} \right ] ,
\end{equation}
where the splitting functions $f_{MB/H}(y)$ and
$f_{BM/H}(y)$ are related to the probability amplitudes $\phi_{BM}$ 
by 
\begin{eqnarray} 
   f_{BM/H}(y)&=&\int_0^\infty dk_\perp^2 |\phi_{BM} (y,k_\perp^2)|^2,  
     \nonumber\\
   f_{MB/H}(y)&=&\int_0^\infty dk_\perp^2 |\phi_{BM} (1-y,k_\perp^2)|^2. 
\end{eqnarray}
They can be calculated by using
time-ordered perturbation theory in the infinite momentum 
frame.  The quark distributions in a physical hadron, $H$  
are then given by
\begin{equation}
     q_H (x) = Z 
q_H^{bare} + \delta q_H(x)  
\end{equation}
where $q_H^{bare}$ are the bare quark distributions 
and $Z$ is a renormalization constant which can be expressed 
as 
\begin{equation}
  Z\equiv 1 - \sum_{MB}
\int_0^1 f_{MB/H} (y) dy.
\end{equation}
These concepts can be  extended to polarized particles 
by introducing the probability amplitudes 
$\phi_{BM}^{\lambda\lambda^\prime} (y,k_\perp )$ 
for a hadron with given positive helicity  to be 
in a Fock state consisting a baryon with helicity $\lambda$ and 
meson with helicity $\lambda^\prime$. The 
splitting functions are then  given by 
\begin{eqnarray} 
   f^\lambda_{BM/H}(y)& = & \sum_{\lambda^\prime} \int_0^\infty 
  dk_\perp^2 |\phi_{BM}^{\lambda\lambda^\prime} 
 (y,k_\perp^2)|^2,  \nonumber\\
 f^{\lambda^\prime}_{MB/H}(y)&=& \sum_{\lambda}
  \int_0^\infty  dk_\perp^2 |\phi_{BM}^{\lambda\lambda^\prime}
  (1-y,k_\perp^2)|^2. 
\end{eqnarray}
The contribution of higher Fock states to 
the polarized quark distributions,  
$\Delta q_H(x) = q_H^\uparrow (x)  -q_H^\downarrow (x)$,   
are then  
\begin{equation}
\Delta\delta q_{H}(x) = \sum_{MB} \left [
  \int_x^1 \Delta f_{BM/H}(y) \Delta q_B (\frac{x}{y})
\frac{dy}{y} 
+    \int_x^1 \Delta f_{MB/H}(y) \Delta q_M (\frac{x}{y})
\frac{dy}{y} \right ],  
\label{polsplit} 
\end{equation} 
where 
$\Delta f_{BM/H} (y)$ 
and  $\Delta f_{BM/H} (y)$ are defined by  
$\Delta f_{BM/H} (y)\equiv \sum_{\lambda} 2\lambda   
f^{\lambda}_{BM/H} (y)$ and 
$\Delta f_{MB/H} (y)\equiv \sum_{\lambda^\prime} 
2\lambda^\prime f^{\lambda^\prime}_{BM/H} (y)$, respectively. 
The contributions from the second term in Eq. (\ref{polsplit})  
are zero for pseudoscalar mesons.  

The amplitudes 
$\phi^{\lambda\lambda^\prime}_{BM}(y,k_\perp^2)$ may be expressed in the 
following form 
\begin{equation}
\phi^{\lambda\lambda^\prime}_{BM}(y,k_\perp^2)
 = \frac{1}{2\pi\sqrt{y(1-y)}} 
  \frac{\sqrt{m_Hm_B}V^{\lambda\lambda^\prime}_{IMF}(y,k_\perp^2)}
      {m^2_H-{\cal M}_{BM}^2(y,k_\perp^2 )}. 
\end{equation}
Here, $V^{\lambda\lambda^\prime}_{IMF(y,k_\perp^2)}$ 
describes the vertex and  
contains  the spin-dependence of the
amplitude. The exact form of the 
$V^{\lambda\lambda^\prime}_{IMF(y,k_\perp^2)}$  
can be found for various transitions in Refs.\cite{SpeT} and \cite{HSS}.  
Because of the extanded nature of the vertices one has to 
introduce phenomenological vertex form factors, 
$G_{HBM}(y,k^2_\perp )$, which parametrize the unknown dynamics at the 
vertices. These  are often parametrized  as  
\begin{equation} 
     G_{HBM}(y,k_\perp^2 )= 
\left( \frac{\Lambda^2_{BM}+ m_H^2}{
\Lambda^2_{BM}+{\cal M}^2_{BM}(y,k_\perp^2)}\right)^2 , 
\end{equation}
where 
\begin{equation}
{\cal  M}_{BM}^2 =\frac{k_\perp^2 + m_B^2}{y} +
\frac{k_\perp^2 + m_M^2}{1-y}
\end{equation}
is the invariant mass of the meson-baryon fluctuation.

In calculating the matrix element of the axial-current or  
$g_1$ in  the meson-cloud model, one has to include terms 
in which the polarized photon-N (photon-$\Sigma$) interaction leads to the 
same final states as the polarized photon-$\Delta$ 
(photon-$\Sigma^*$) interaction \cite{cross}. 
The contributions of these interference terms to the 
measured quark distributions can be written as 
\begin{equation} 
\Delta\delta^{int} q_{H}(x) = \sum_{MB_1B_2} \left [
  \int_x^1 \Delta f_{(B_1B_2)M/H}(y) \Delta q_{B_1B_2} (\frac{x}{y}) 
\frac{dy}{y} 
+    \int_x^1 \Delta f_{(M_1M_2)B/H}(y) 
\Delta q_{M_1M_2} (\frac{x}{y}) \frac{dy}{y} \right ],  
\label{interf} 
\end{equation} 
where the interference splitting functions are given by 
\begin{eqnarray} 
  \Delta f_{(B_1B_2)M/H}(y) &=& \sum_{\lambda\lambda^\prime} 
      2\lambda \int_0^\infty dk_\perp^2 
      \phi_{B_1M}^{\lambda\lambda^\prime}(y,k_\perp^2) 
      \phi_{B_2M}^{*\lambda\lambda^\prime}(y,k_\perp^2) \nonumber\\ 
  \Delta f_{(M_1M_2)B/H}(y) &=& \sum_{\lambda\lambda^\prime} 
      2\lambda^\prime  \int_0^\infty dk_\perp^2 
      \phi_{M_1B}^{\lambda\lambda^\prime}(y,k_\perp^2) 
      \phi_{M_2B}^{*\lambda\lambda^\prime}(y,k_\perp^2).  
\end{eqnarray} 

\begin{figure}
\psfig{figure=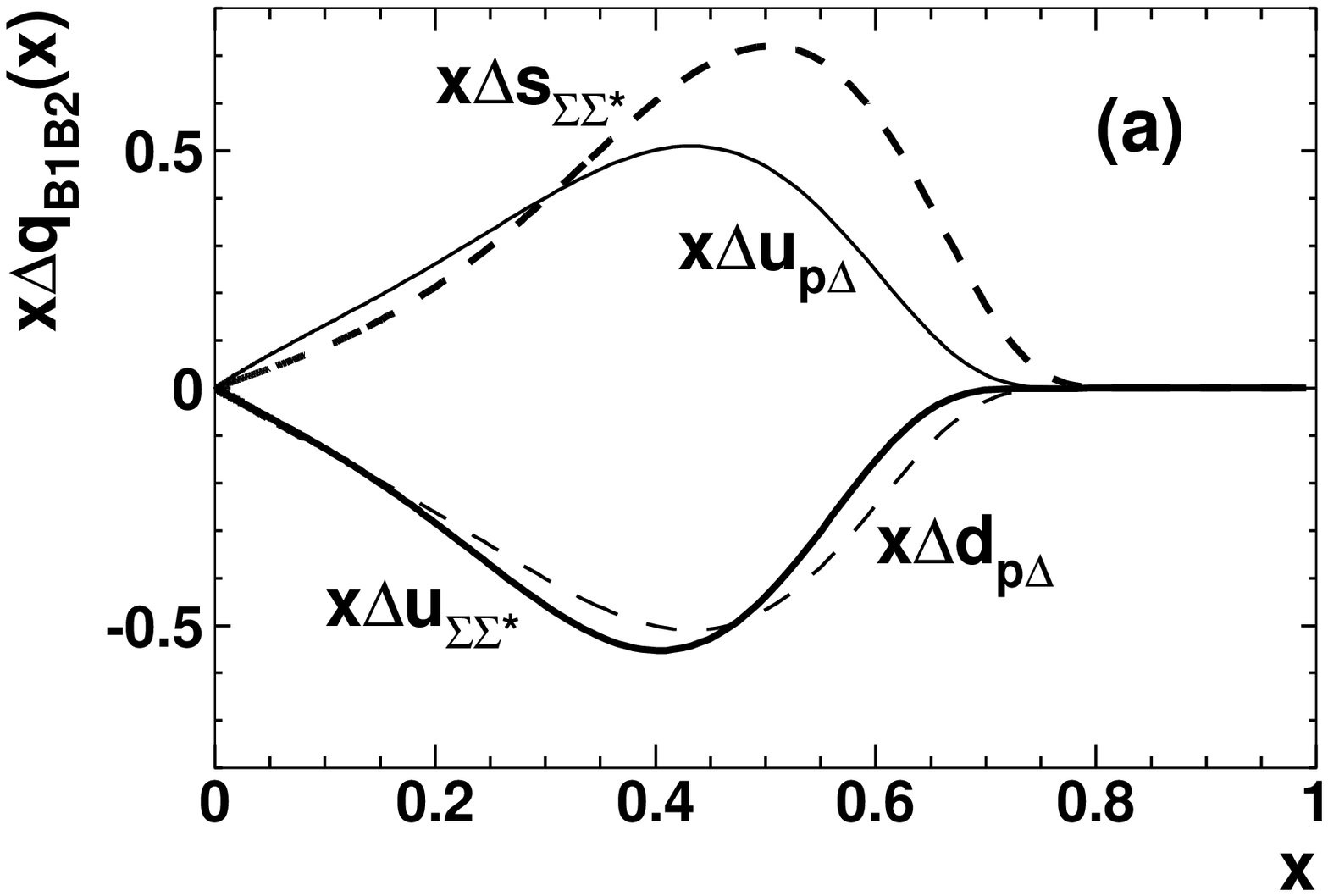,height=8.cm}

\psfig{figure=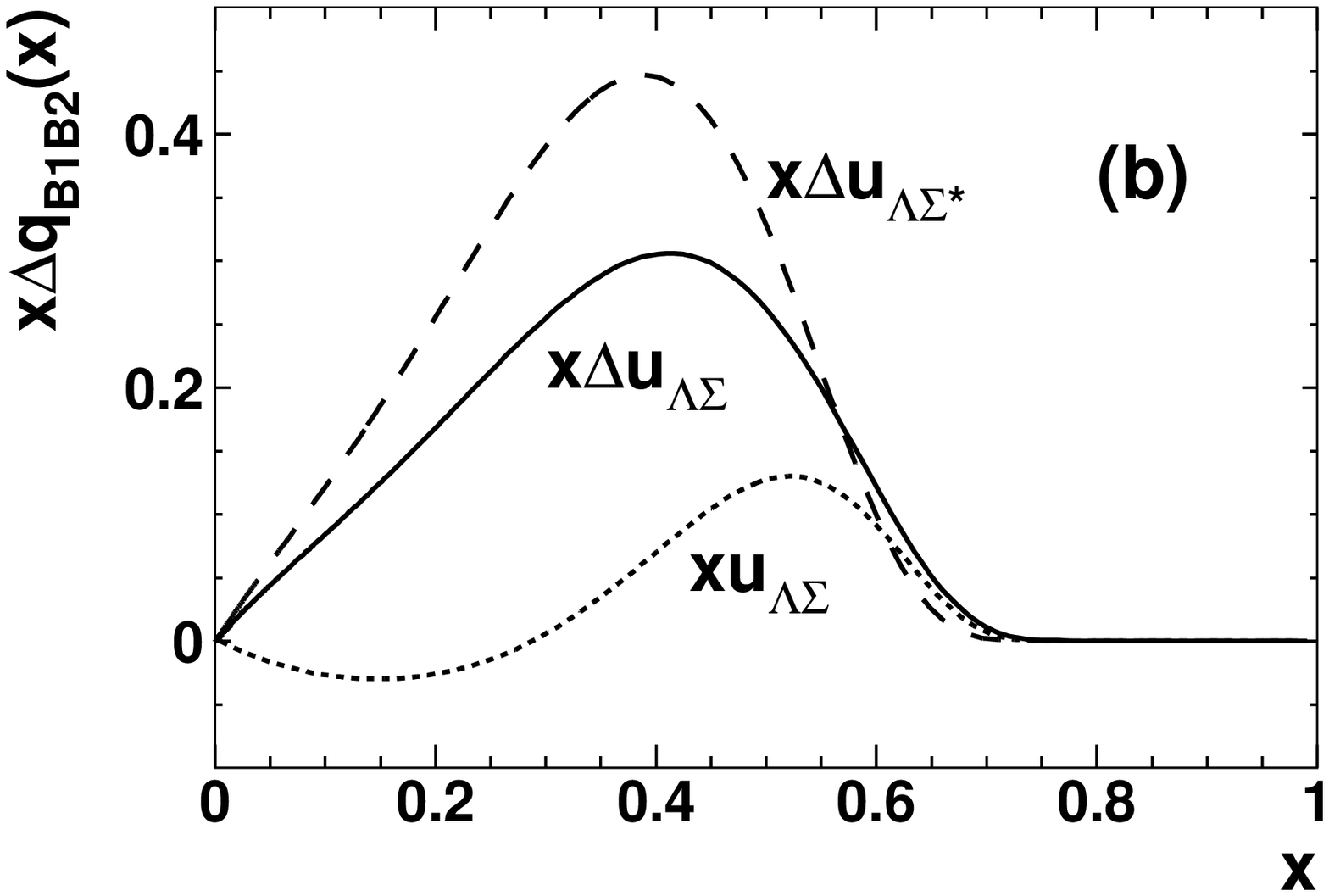,height=8.cm}
\caption{Interference distributions as calculated in the 
MIT bag at the scale, $\mu^2$. 
(a) $N$-$\Delta$  and $\Sigma$-$\Sigma^*$ interference terms; 
(b) $\Lambda$-$\Sigma$ and $\Lambda$-$\Sigma^*$ intereference terms.  
The $d$ distributions have the same magnitude but opposite signs than 
the corresponding $u$-distributions. }  
\label{fig10} 
\end{figure}

The interference  distributions $q_{B_1B_2}$ and 
$q_{M_1M_2}$ in Eq. (\ref{interf}) do not  
have the same straightforward interpretation as 
quark distributions. They have to be modeled in some way. 
Using the $SU(6)$ wave functions of the baryons from the 
baryon octet and decuplet, the transition matrix elements,  
$\langle B_8 | P_{f,m}| B_{8}^\prime \rangle$ and 
$\langle B_8 | P_{f,m}| B_{10} \rangle$, 
may be calculated 
and the interference distributions may be related to 
the quatities $F_s$, $F_v$, $G_v$ and $G_s$ 
calculated in the MIT bag. For the 
$\Delta N$ interference terms we obtain  
\begin{eqnarray} 
  u_{\Delta^+ p}^{\uparrow\downarrow} & = & 
  u_{\Delta^0 n}^{\uparrow\downarrow} 
  = \pm \frac{\sqrt{2}}{3} G_v(x) \nonumber \\
  d_{\Delta^+ p}^{\uparrow\downarrow} & = & 
  d_{\Delta^0 n}^{\uparrow\downarrow} 
  = \mp \frac{\sqrt{2}}{3} G_v(x).  
\end{eqnarray}
The possible interference terms for the 
$\Sigma^+$ are 
$\Lambda$ -$\Sigma^0$, $\Lambda$ -$\Sigma^{*0}$, 
$\Sigma^0$ -$\Sigma^{*0}$ and $\Sigma^+$ -$\Sigma^{*+}$.  
We obtain for the $\Lambda$-$\Sigma^0$ interference  
\begin{eqnarray} 
  u_{\Lambda \Sigma^0}^{\uparrow\downarrow} & = & 
  - d_{\Lambda \Sigma^0}^{\uparrow\downarrow} = 
    \frac{\sqrt{3}}{8} [F^\prime_s-F^\prime_v] 
    \pm \frac{\sqrt{3}}{8} 
     [G^\prime_s+\frac{1}{3}G^\prime_v] \nonumber\\ 
  s_{\Lambda \Sigma^0}^{\uparrow\downarrow} & = & 0,  
\end{eqnarray} 
for the $\Lambda$-$\Sigma^{*0}$ interference 
\begin{eqnarray} 
  u_{\Lambda \Sigma^{*0}}^{\uparrow\downarrow} & = & 
  - d_{\Lambda \Sigma^{*0}}^{\uparrow\downarrow} =
  \pm \frac{1}{\sqrt{6}} G^\prime_v(x) \nonumber \\
  s_{\Lambda \Sigma^{*0}}^{\uparrow\downarrow} & = & 0,  
\end{eqnarray}
and for the $\Sigma^0$-$\Sigma^{*0}$ 
and $\Sigma^+$-$\Sigma^{*+}$ interference terms 
\begin{eqnarray} 
  u_{\Sigma^{*+} \Sigma^+}^{\uparrow\downarrow} & = & 
  2 u_{\Sigma^{*0} \Sigma^0}^{\uparrow\downarrow} = 
  2 d_{\Sigma^{*0} \Sigma^0}^{\uparrow\downarrow}  =  
  \mp \frac{\sqrt{2}}{3} G^\prime_v(x) \nonumber \\
  s_{\Sigma^{*+} \Sigma^+}^{\uparrow\downarrow} & = & 
  s_{\Sigma^{*0}\Sigma^0}^{\uparrow\downarrow} 
  = \pm \frac{\sqrt{2}}{3} G_v(x) . 
\end{eqnarray}
We use the average mass of the octet and decuplet baryons 
involved in the calculation of  $F_s$, $F_v$, $G_s$ and $G_v$.  
The results for  $p$-$\Delta$ and $\Sigma$-$\Sigma^*$  
are shown in Fig. \ref{fig10}a and those for $\Lambda$-$\Sigma^*$ 
and $\Lambda$-$\Sigma^*$ in Fig. \ref{fig10}b. The 
$d$ distributions for the $\Lambda$-$\Sigma$ interference terms 
can be obtained  
by multiplying  the corresponding $u$ distributions by $-1$. 
Note that, if $SU(6)$ is not a good symmetry, we have $F_s\ne F_v$ and 
the $\Lambda$-$\Sigma^0$ interference also 
contributes  to the the unpolarized $u$ and $d$ quark 
distributions. This is shown in Fig. \ref{fig10}b. 
However, the net contributions, i.e. the integral over 
$u_{\Lambda\Sigma^0}$ and that over $d_{\Lambda\Sigma^0}$,  
are  zero and baryon number conservation is not 
violated. Note also that the interference distributions 
for $\Sigma$-$\Sigma^*$ 
and $N$-$\Delta$  have opposite signs. 
Nevertheless, they contribute both positively to $\Delta u$ 
since the splitting functions have opposite signs as we shall 
discuss below.

In order to calculate the meson cloud corrections 
to the quark distributions we have to specify the 
coupling constants and the cut-off parameters.  
SU(3) relates the  coupling constants by 
$g_{\Sigma\Sigma\pi} = 2(1-\alpha) g_{NN\pi}$,  
$g_{\Sigma\Lambda\pi}= \frac{2}{\sqrt{3}} \alpha g_{NN\pi}$, 
$g^2_{\Sigma^+p \bar K^0}  =  \sqrt{2}(1-2\alpha) g_{NN\pi}$ and 
$g_{\Sigma\Sigma^*\pi} =\frac{1}{\sqrt{6}}
g_{N\Delta\pi}$ where we defined 
$g_{NN\pi}\equiv g_{pp\pi^0}$,    
$g_{N\Delta\pi}\equiv  g_{p\Delta^{++}\pi^-}$,    
$g_{\Sigma\Sigma\pi}\equiv g_{\Sigma^+\Sigma^+\pi^0}$,  and 
$g_{\Sigma\Sigma^*\pi}\equiv g_{\Sigma^+\Sigma^{*+}\pi^0}$. 
$\alpha$ is defined by       
$\alpha\equiv D/(D+F)\approx 0.635$ with $D$ and $F$ the 
symmetric and antisymmetric SU(3) couplings.   
The numerical values are given by 
$g^2_{pN\pi}/4\pi=13.6$ and $g^2_{p\Delta\pi}/4\pi = 11.08$ GeV$^{-2}$ 
and the couplings of a given type of fluctuation with different 
isospin components are related by isospin Clebsch-Gordon coefficients, 
i.e. $g_{pn\pi^+}=- \sqrt{2}g_{pp\pi^0}$, 
$g_{p\Delta^0\pi^+}=
-\frac{1}{\sqrt{2}}g_{p\Delta^+\pi^0}= 
\frac{1}{\sqrt{3}}g_{p\Delta^{++}\pi^-}$,   
$g_{\Sigma^+\Sigma^0\pi^+}=-g_{\Sigma^+\Sigma^+\pi^0}$,  
and  
$g_{\Sigma^+\Sigma^{*0}\pi^+}=-g_{\Sigma^+\Sigma^{*+}\pi^0}$. 
The cut-off parameters may be determined in independent 
experiments, for example in  inclusive particle production 
in hadron hadron collisions \cite{Zoller,SpeT,Boros}. 
The violation of the Gottfried sum rule and of flavor symmetry 
puts also constraints on the magnitude of these parameters. 
They are also restricted by the requirement that 
the contributions from the meson cloud to the 
sea quark distributions cannot be larger than the measured sea quark 
distributions. The values, $\Lambda_{MB} =1.0$ GeV 
and $\Lambda_{MB} =1.3$ GeV  
for the $\pi N$ and $\pi\Delta$ components, respectively,  
give contributions to the $\bar u$ and $\bar d$ which  
are consistent with this requirement and 
also with FSV violation \cite{MT98} (see below). 
Unfortunately, there is not much known about the cut-off 
parameters in the $\Sigma^+$ case. In the absence of any information, 
we use the same values as in the proton case. 
With this choice of parameters  
the probabilities for the various fluctuations are approximately  given 
by $P_{N\pi/p}=13\%$, $P_{\Delta\pi/p}=11\%$ and 
$P_{\Sigma\pi/\Sigma}=3.7\%$, $P_{\Sigma^*\pi/\Sigma}=3.1\%$, 
$P_{\Lambda\pi/\Sigma}=3.2\%$  and $P_{p\bar K^0/\Sigma}=0.4 \%$,  
respectively.   

\begin{figure}
\psfig{figure=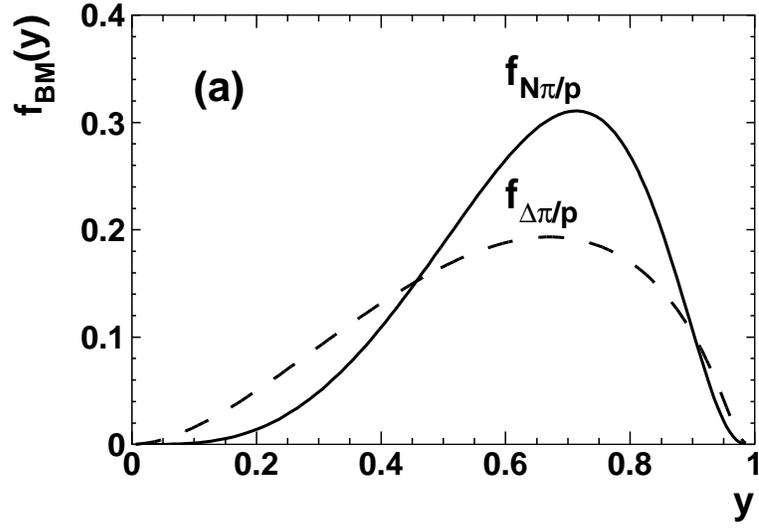,height=8.cm}

\psfig{figure=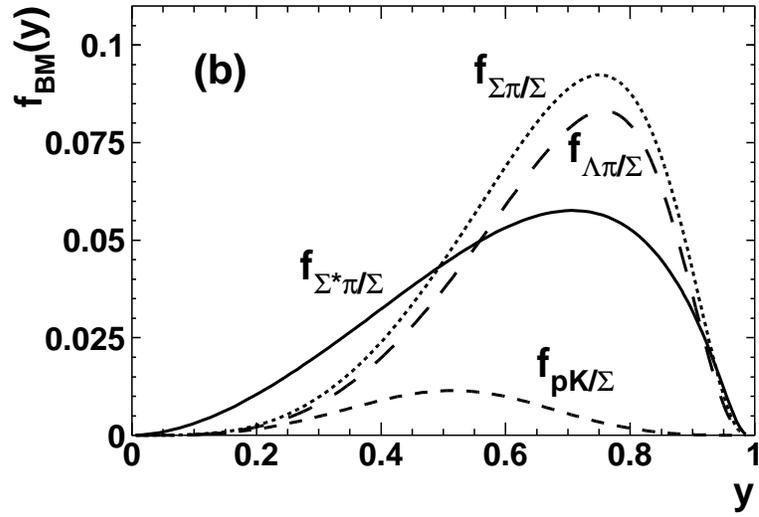,height=8.cm}
\caption{Splitting functions  
for the transitions (a)  \protect$p\rightarrow BM$   
 and (b) \protect$\Sigma^+\rightarrow BM$.} 
\label{fig11}
\end{figure}

The spin averaged splitting functions for $p\rightarrow BM$ and 
$\Sigma^+ \rightarrow BM$ 
are shown in Fig. \ref{fig11}.  
Here, the splitting functions for a given type of  fluctuation 
are defined as the sum over all isospin states -- i.e.,  
$f_{N\pi/p}\equiv f_{p\pi^0/p}+f_{n\pi^+/p}$, etc. 
Because of  the smaller coupling constants in the $\Sigma^+$ case 
the meson-cloud is less important for the $\Sigma^+$. 
Further, the transition, $\Sigma^+ \rightarrow \bar K^0 p$, 
only plays a marginal role as can be seen in Fig. \ref{fig11}b. 
In calculating the meson-cloud corrections we use 
our bag model results for the bare distributions of the 
hyperons and nucleons. We also use a parametrization of the 
quark distributions in the pions \cite{PionStr} and utilize 
experimental data for the ratio $\bar u^{K^-}/\bar u^{\pi^-} 
\sim (1-x)^{0.18\pm0.07}$ \cite{Kaon} 
to obtain the light quark valence distribution 
in the kaon. The strange quark distribution in the kaon 
is expected to be harder because of the mass of the strange quark. 
We use the parametrization of Ref.\cite{Alb98} for 
the kaon quark distributions,  
which are constructed to fullfill the above requirements:  
\begin{eqnarray} 
 xu(x) & = & 1.05 x^{0.61} (1-x)^{1.20},  \nonumber \\ 
 xs(x) &= & 0.94x^{0.61} (1-x)^{0.86}.  
\end{eqnarray} 

\begin{figure}
\psfig{figure=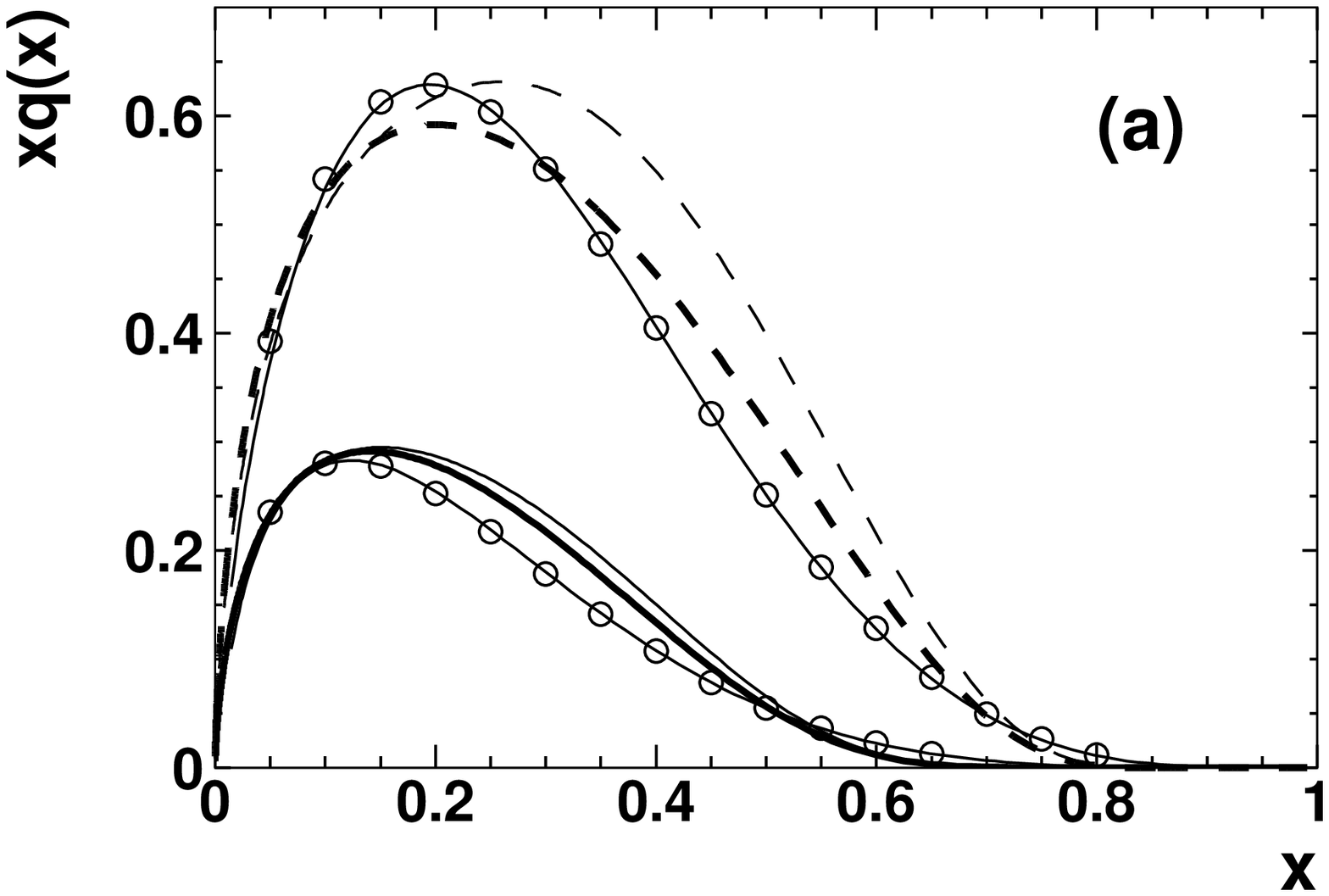,height=8.cm}

\psfig{figure=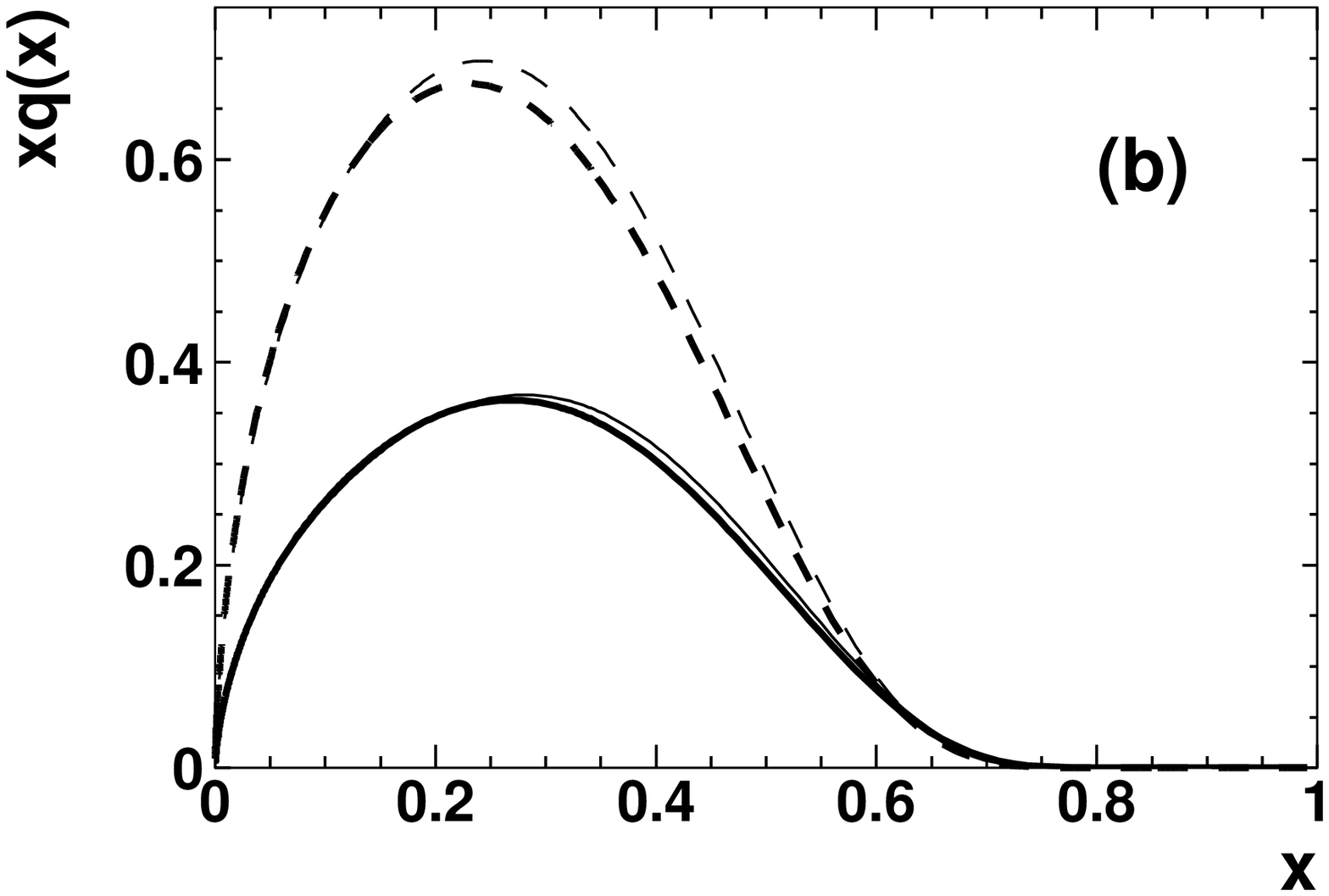,height=8.cm}
\caption{(a) The up (dashed lines) and down 
(solid lines) valence quark distribution in the
proton without (light lines)
and with  (heavy lines) 
meson-cloud corrections at $Q^2=10$ GeV$^2$. 
The Cteq4M distributions representing the ``data''  
are shown as solid lines with open circles. 
(b) The up (dashed lines) and strange  
(solid lines) valence quark distribution in the
\protect$\Sigma^+$  without (light lines)
and with  (heavy lines) 
meson-cloud corrections at $Q^2=10$ GeV$^2$. } 
\label{fig12}
\end{figure}

First, we show the modifications of the bare  valence  quark distributions 
in the proton and in the $\Sigma^+$ in Fig. \ref{fig12}.  
We see that the meson-cloud plays a relatively  
more important role in the proton than in the $\Sigma^+$. 
The strange to light quark ratio, 
$r_\Sigma = s_\Sigma/u_\Sigma$, is not sensitive to meson-cloud 
corrections, as shown in Fig.\ref{fig4}. 

\begin{figure}
\psfig{figure=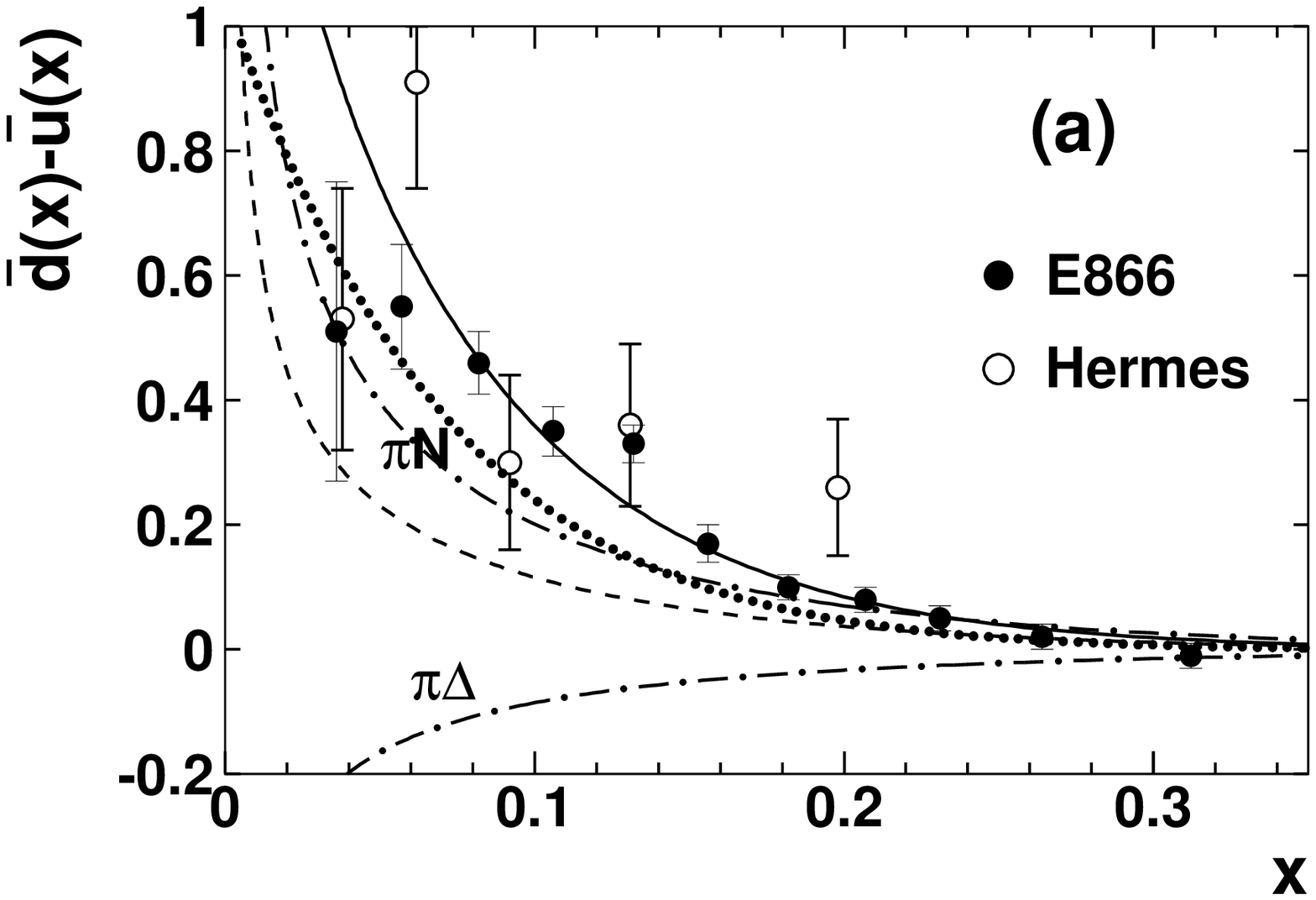,height=8.cm}

\psfig{figure=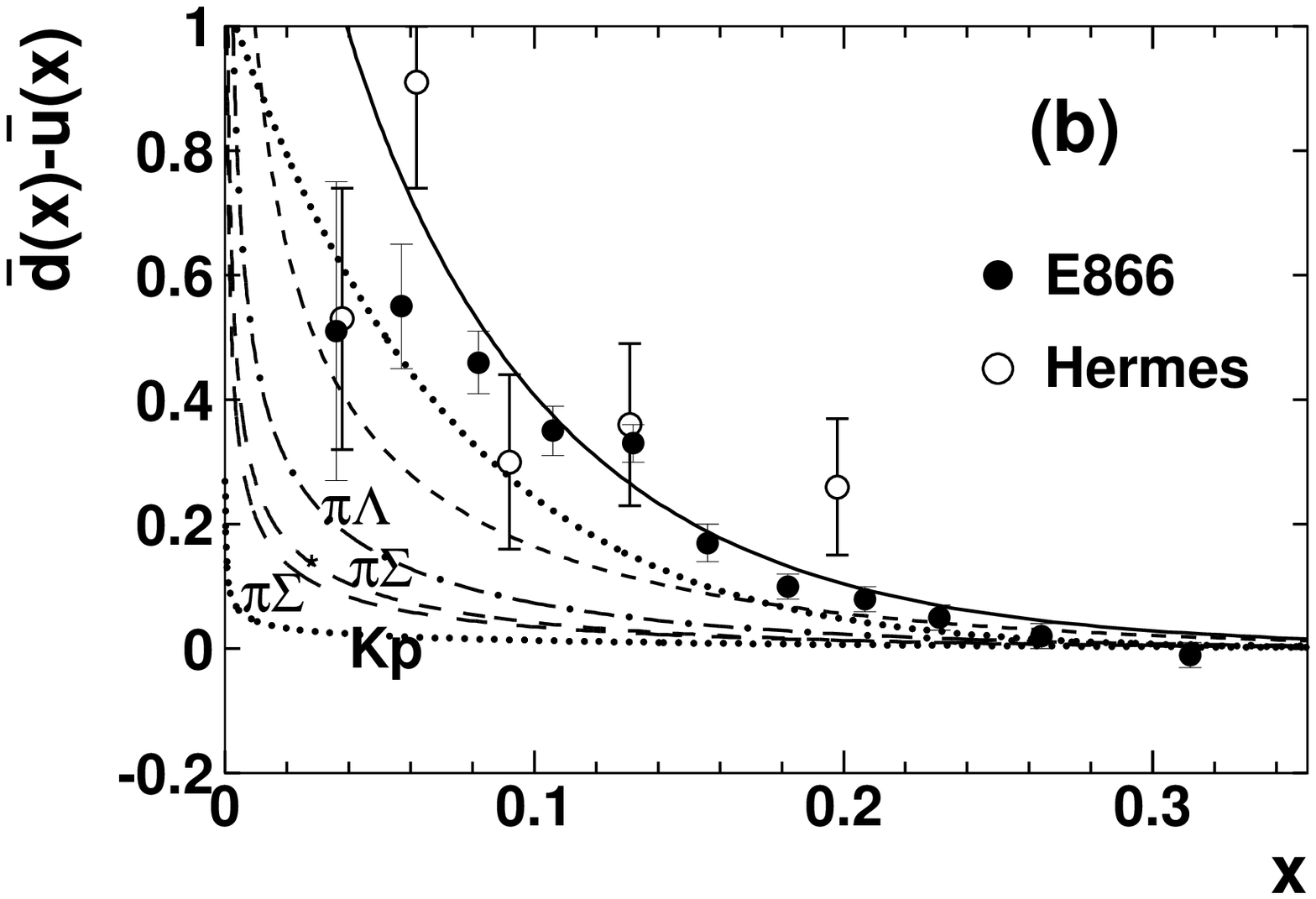,height=8.cm}
\caption{Flavor symmetry violation, 
$\bar d -\bar u$, for (a) the proton and (b) the  $\Sigma^+$.   
In the proton case,  
the upper and lower dash-dotted lines stand for the 
$\pi N$ and $\pi\Delta$ contributions 
alone and the dashed line for their sum. 
The data are  taken 
from Refs. \protect\cite{E866,HER}. 
In the $\Sigma^+$ case, 
the upper and lower dashed lines stand for the 
$\pi\Sigma$ and 
$\pi\Sigma^*$ contributions, the dotted line 
for the $Kp$  and the dash-dotted line 
for the $\pi\Lambda$ contributions -- and the proton data is shown just
to set the scale.  
The short dashed line is the sum of the 
chiral components. The dotted lines are 
the Pauli contributions and the solid lines stand for the total 
FSV.  }  
\label{fig13}
\end{figure}

The meson cloud model predicts  flavor symmetry violations 
not only for the proton but also for other baryons. 
Since $p\leftrightarrow \Sigma^+$ means 
$d(\bar d) \leftrightarrow s(\bar s)$ under $SU(3)$, one would expect 
an excess of $\bar s$ over $\bar u$ on the basis of complete 
SU(3) symmetry and the measured FSV in the proton.  
However, in the meson-cloud model, 
$s$-$\bar s$ fluctuations for the $\Sigma^+$ 
involve hyperons containing at least two 
strange quarks, $\Xi$'s, and are strongly  suppressed due to the 
higher masses of these hyperons, which is of course 
a direct consequence of SU(3) breaking. 
On the other hand,   
meson-cloud contributions lead to an excess of $\bar d$ over 
$\bar u$ for the $\Sigma^+$, as can be seen in Eq.(\ref{sfluct}). 
This FSV is not at all related to 
$SU(3)$ symmetry.  Furthermore,  
FSV could be even larger in the $\Sigma^+$ case since here {\it all} 
fluctuations contribute to $\bar d$. 
We show the calculated FSV violation 
for the proton in Fig.\ref{fig13}a and 
for the $\Sigma^+$ in Fig.\ref{fig13}b, together with the 
E866 data for the proton. For the proton, 
the upper and lower dash-dotted curves are  the 
contributions from the $\pi N$ and the $\pi\Delta$ components alone 
and the dashed curve is the sum of $\pi N$ and $\pi\Delta$.  

As pointed out in Ref. \cite{MT98}, 
the measured $x$-dependence of the FSV, especially 
that of the ratio $\bar d/\bar u$ (not shown), requires a relatively  
large contribution from the $\Delta\pi$ component in the proton 
case, which cancels the contributions from the $N\pi$  
component at large $x$ values and leads to the required 
fast decrease of the asymmetry in this region.   
Since, on the other hand, the magnitude of the 
$N\pi$ and $\Delta\pi$ components are restricted by 
the requirement that their contributions to the 
total sea quark distributions can not larger than the experimentally  
measured value, an additional non-chiral component 
is needed  at small $x$. This non-chiral component may be attributed 
to the Pauli exclusion principle, as suggested by Field and 
Feynman \cite{Field}. Because of the Pauli exclusion principle, 
the presence of two valence $u$ quarks in the proton, 
as opposed to a single valence $d$ quark, makes 
it less probable to produce a $u\bar u$ pair 
compared to a $d\bar d$ pair giving an excess of 
$\bar d$ over $\bar u$ in the non-perturbative sea.  
Based on bag model calculations \cite{Schreiber}, 
it is expected that this component should have a shape similar 
to the usual sea quark distributions, contributing 
to the asymmetry at lower $x$ values 
than the chiral component. Since we 
have  two valence $u$ quarks and {\it no} $d$ valence quarks
in the $\Sigma^+$ we expect that the component arising from  
the Pauli principle will be at least as large as that in the proton 
case. (Neglecting SU(3) breaking one would expect it to be twice as large 
for the $\Sigma^+$ as for the proton.) 
The Pauli contributions are shown as the dotted 
line and the sum of the chiral and Pauli components 
as solid lines in Figs. \ref{fig13}a and \ref{fig13}b.  
In the $\Sigma^+$ case, 
we show also the  contributions from the various 
meson baryon fluctuations,  
the upper and lower dashed curves stand for  the 
$\pi\Sigma$ and $\pi\Sigma^*$  contributions, 
the dotted line for the $Kp$ and the dash-dotted for 
the $\pi\Lambda$ contribution. 
The sum of all chiral contributions is shown as 
the short dashed line. 
Note that,  
while the contribution of the 
$\pi\Delta$ component is negative in the 
proton case, the $\pi\Sigma^*$ component reinforces the FSV 
in the $\Sigma^+$ case giving rise to as large a FSV as in the 
proton case --- even  though the total meson-cloud corrections  
are less important for the $\Sigma^+$. 

\begin{figure}
\psfig{figure=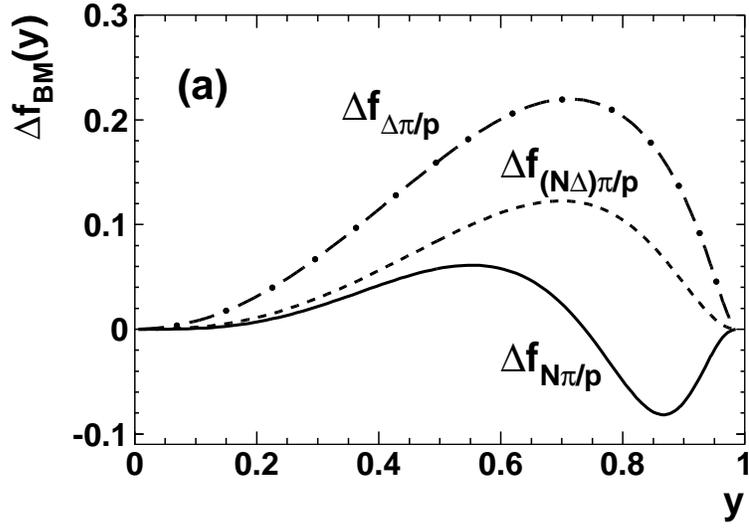,height=8.cm}

\psfig{figure=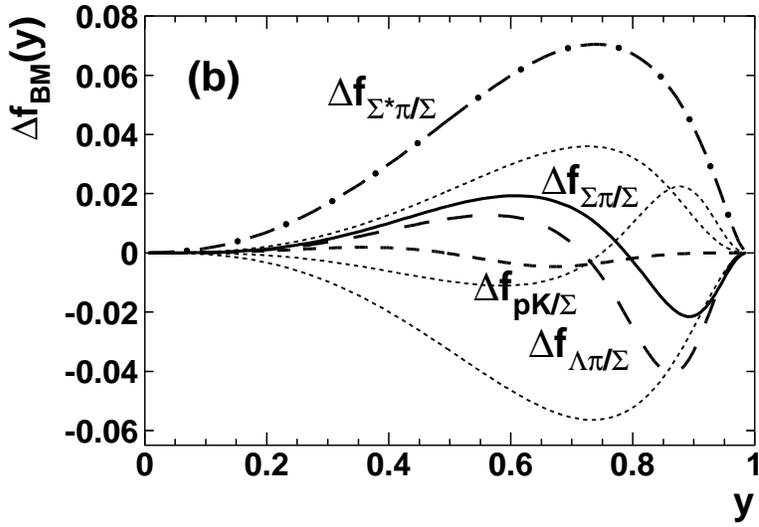,height=8.cm}
\caption{The polarized splitting functions 
for the transitions (a) 
$p\rightarrow BM$  and (b) $\Sigma^+\rightarrow BM$. 
 The dotted lines are the interference splitting functions; 
$\Delta f_{(\Lambda\Sigma^*)\pi/\Sigma}$ (upper line), 
$\Delta f_{(\Lambda\Sigma)\pi/\Sigma}$ (middle line) 
and $\Delta f_{(\Sigma\Sigma^*)\pi/\Sigma}$ 
(lower line). }  
\label{fig14}
\end{figure}

Since the pseudoscalar mesons do not contribute  to the  
spin dependent quark distributions  of the baryons,  
the meson-cloud corrections decrease the amount of the baryon 
spin carried by the spin of the quarks.   
The polarized splitting functions for the 
proton and $\Sigma^+$ are shown in Figs. \ref{fig14}a and \ref{fig14}b,  
respectively. (Note that, according to the definition, 
the decuplet splitting functions are the sum of the $3/2$ and 
$1/2$ helicity components with the $3/2$ component multiplied 
by a factor of $3$.)  
Since the fluctuations involving baryons from the octet are positive 
for small $y$ values and negative for larger $y$ values,  their 
contribution to the spin of the nucleon or hyperon is relatively small. 
The integral,  $\langle \Delta f_{N\pi/p}\rangle \approx 0.01$,   
nearly vanishes.  
On the other hand, the splitting functions of the baryons from the 
baryon decuplet are positive over the whole $y$ region and their 
contributions is much larger, 
$\langle \Delta f_{\Delta\pi/p}\rangle \approx 0.11$.  
These values  are to be compared to 
the values $\langle  f_{N\pi/p}\rangle =0.13$ and  
$\langle  f_{\Delta\pi/p}\rangle =0.11$ which can be roughly thought of 
as the amount of spin ``lost'' through the meson baryon fluctuation. 
(Remember that mesons do not contribute to the spin of the 
nucleon.)  
The splitting functions 
corresponding to interference  
between octet and decuplet baryons 
(short dashed lines) are positive. However, since the 
interference distributions for $d$ and $s$ quarks are 
opposite in sign to the $u$ distributions (see 
Fig. \ref{fig10}) they approximately cancel each other in 
the ``spin sum''. On the other hand, they contribute positively 
to $g_1$ since, here, the $u$ distributions are weighted 
by $4/9$ as opposed to $1/9$ of the 
$d$ and $s$ distributions.

In Table 1, we show spin fractions carried by the different 
flavors of the proton and the $\Sigma^+$,  
$\Delta Q\equiv \Delta q+\Delta\bar q$, in the 
non-relativistic quark model (NQM), as measured in DIS and 
using SU(3) symmetry to obtain the values for $\Sigma^+$ 
(DIS+SU(3)), in the bag model, in the bag model 
with meson cloud corrections (Bag+MC) and 
with interference terms (IF). $\Delta S$ in the proton 
comes from the $\Lambda K$ and $\Sigma K$ component 
of the wave function. However, these give  very small 
contributions. $\Delta D$ in the $\Sigma^+$ comes from 
lower lying fluctuations and could be sizable. 
However, because the integral over the splitting function 
for the octet baryons approximately vanishes, it is very small 
($< 1 \%$).  Here, the interference terms largely  cancel 
each other.  
Further, we see that,  
because of the transverse motion of the quarks, 
the fraction of the spin  carried 
by the quarks in the bag model is smaller than one. 
In conclusion, the meson cloud is responsible for part of the dilution 
of the spin though the fraction of spin carried by the 
quarks is still considerable larger than the experimental value.  

\begin{figure}
\psfig{figure=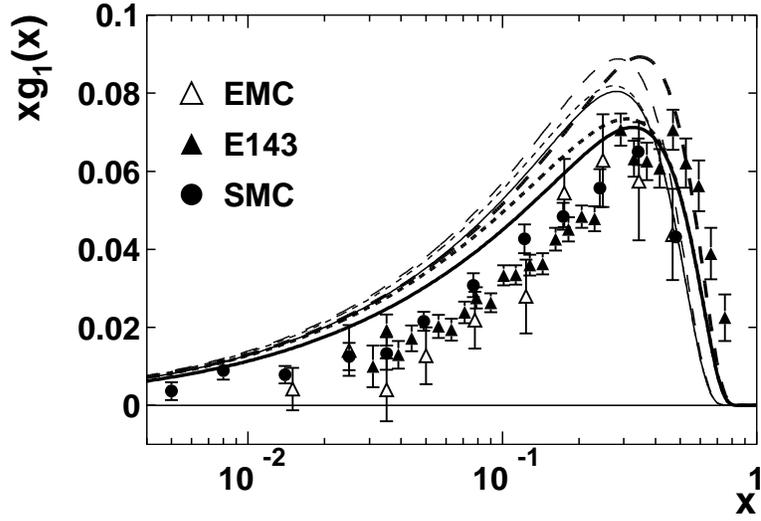,height=8.cm}
\caption{\protect $g_{1p}$ 
with (light solid line) and  without (light dashed line) 
meson-cloud corrections compared to the 
corresponding $g_{1\Sigma^+}$ in the proton (heavy lines). 
The data are for the proton and taken 
 from Refs. \protect\cite{EMC,SMC,E143}. 
The structure functions calculated with interference terms are shown 
as short dashed lines. 
The EMC data are at different $Q^2$ values, the 
SMC at $Q^2=10$ GeV$^2$ and the E143 data at $Q^2=5$ GeV$^2$.} 
\label{fig15}
\end{figure}

In Fig. \ref{fig15}, $xg_1(x)$ calculated for proton
(heavy lines) and for $\Sigma^+$ (light lines) are shown
with (solid lines) and without (dashed lines) meson corrections
and with interference terms (short dashed lines).
The predictions for $g_{1\Sigma^+}$ and
$g_{1p}$ are similar, with $g_{1\Sigma^+}$ peaking
at slightly lower $x$-values
than $g_{1p}$. This is because the $u$ quarks,
which have a somewhat softer distribution in the $\Sigma^+$ than
in the proton, dominate in $g_1$.

In concluding this section we must issue a caution concerning the 
discussion of spin-dependent parton distributions here. It is 
by now well understood that the axial anomaly plays a vital role 
in the flavor singlet spin structure \cite{ANOM} and the model which we have 
used has not incorporated such effects. As a result the integral of 
$g_{1p}$, for example, satisfies the Ellis-Jaffe sum rule -- with 
the octet and isovector axial charges appropriate to the model, including 
meson corrections. It is therefore not too surprising that our 
curves for $g_{1p}$ lie above the data.
A reasonable polarized gluon distribution could bring 
the calculated values in Table. 1 into better agreement with 
the experimental value of $\Sigma$.   

\section{Conclusions} 

We calculated the quark distribution functions of different hyperons 
in the MIT bag model using the approach of the Adelaide group which 
assures the correct support of the distribution functions.  
The hyperfine splitting responsible for the 
splitting of the masses of the $N$-$\Delta$, $\Lambda$-$\Sigma^0$ and 
$\Sigma$-$\Sigma^*$ results in quark distributions very different from 
$SU(6)$ expectations. 
This $SU(6)$ breaking goes beyond the 
explicit breaking through the strange quark mass and leads to 
different shapes of the quark distributions, even in hyperons 
with the same number of (valence) strange quarks. 
The strange to u ratio in the $\Sigma^+$ increases with 
$x\rightarrow 1$ --- a behaviour opposite to that  
predicted by $SU(3)$. 
Further, we predict polarized $u$ and $d$ quarks distributions 
in the $\Lambda$  as a function of $x$, even though their net 
contributions to the total spin of the $\Lambda$ are zero. 
This prediction could be tested in semi-inclusive polarized 
DIS since the 
coupling of the $u$ quarks to the electromagnetic current 
is four times larger than 
that of  the strange quarks. 

We also calculated the modifications of the bare quark distributions 
through the meson-cloud required by chiral symmetry. 
Although the meson-cloud corrections to the distributions 
in the $\Sigma^+$ are not as large as those to the corresponding 
distributions in the proton, because of the smaller coupling 
constants,  
the meson-cloud also leads to significant flavor 
symmetry violations in the sea quark distribution  
of the hyperons. 
We found that the $\bar d$ in the $\Sigma^+$ is enhanced relative to the 
$\bar u$, contrary to $SU(3)$ expectations.  
The $\bar d_{\Sigma^+}/\bar u_{\Sigma^+}$ ratio  
is comparable to the corresponding ratio  
$\bar d_p/\bar u_p$ in the proton since, in the $\Sigma^+$ case, 
all of the lowest lying fluctuations enhance the $\bar d$ relative 
to $\bar u$.

\acknowledgements 

We would like to thank S. Braendler, W. Melnitchouk,   
A.W. Schreiber,  F. Steffens and K. Tsushima for 
many useful conversations. This work was partly supported 
by the Australian Research Council.

\begin{table}
\caption{Fraction of angular momentum  carried by the spin of
the quarks in different models. $\Delta Q \equiv \Delta q + 
\Delta\bar q$. The results in the third row
are obtained by using $\Sigma=0.28$ from deep inelastic scattering
(DIS) experiments, $F+D=1.2573$ and $F/D=0.575$ from hyperon
decay experiments.}

\begin{tabular}{||l||c|c|c|c||c|c|c|c||}
\rule[-0.4cm]{0cm}{1cm}
& \multicolumn{4}{c||}{proton} &
\multicolumn{4}{c||}{$\Sigma^+$}
\rule[-0.4cm]{0cm}{1cm}\\
\hline
\hline
\rule[-0.4cm]{0cm}{1cm}
Model  & $\Delta U$ & $\Delta D$ & $\Delta S$  & $\Sigma$
 & $\Delta U$ & $\Delta D$ & $\Delta S$  & $\Sigma$
\rule[-0.4cm]{0cm}{1cm}\\
\hline
\hline
NQM & $4/3$ & $-1/3$ & $0$ & $1$ &
$4/3$ & $0$ & $-1/3$ & $1$
\rule[-0.4cm]{0cm}{1cm}\\
\hline
DIS + SU(3) & $0.72$ & $-0.44$ & $-0.10$ & $0.28$
& $0.72$ & $-0.10$ & $-0.44$ & $0.28$
\rule[-0.4cm]{0cm}{1cm}\\
\hline 
Bag & $1.05$ & $-0.26$ & $0$ & $0.79$   & $1.05$
 &  $0$ & $-0.27$ &  $0.78$
\rule[-0.4cm]{0cm}{1cm}\\
\hline
Bag + MC  & $0.86$ & $-0.17$ & $<0.01$ & $0.69$
&$0.93$ & $< 0.01$ & $-0.24$ & $0.69$
\rule[-0.4cm]{0cm}{1cm}\\
\hline
Bag + MC + IF  & $0.94$ & $-0.25$ & $<0.01$ & $0.69$
&$0.98$  & $<0.01$ & $-0.28$  & $0.70$
\rule[-0.4cm]{0cm}{1cm}\\
\end{tabular}
\end{table}

\end{document}